\documentclass{llncs}

\usepackage{amsmath}
\usepackage{amsfonts}
\usepackage{amssymb}

\usepackage{times}
\usepackage{fullpage}
\usepackage{graphicx}
\usepackage{epsfig}
\usepackage{setspace}
\usepackage{subfigure}
\usepackage{alltt}
\usepackage{cite}
\usepackage{ifthen}
\usepackage{paralist}
\usepackage{comment}
\usepackage{color}
\usepackage{url}
\usepackage{fancyvrb}
\usepackage{xspace}

\usepackage{float}
\usepackage{listings}
\usepackage[usenames,dvipsnames,svgnames,table]{xcolor}

\usepackage{bibspacing}
\usepackage{multirow}
\usepackage{rotating}
\usepackage{array}

\usepackage{algorithm}
\usepackage{algorithmicx}
\usepackage{algpseudocode}

\algdef{SE}[DOWHILE]{Do}{doWhile}{\algorithmicdo}[1]{\algorithmicwhile\ #1}%

\algrenewcommand\algorithmicindent{1.0em}%

\algblockdefx[Name]{Struct}{End}%
  [1][Unknown]{\textbf{struct} #1 }%
  {\textbf{end struct} }%

\algnewcommand{\LineComment}[1]{\State \(\triangleright\) #1}

\newcommand{\eg}{\emph{e.g.,}\xspace}
\newcommand{\ie}{\emph{i.e.,}\xspace}

\newcommand{\fref}[1]{Figure~\ref{#1}\xspace}
\newcommand{\sref}[1]{Section~\ref{#1}\xspace}

\newcommand{\fov}{{FoV}\xspace}
\newcommand{\fovs}{{FoVs}\xspace}

\newcommand{\degree}{\ensuremath{^\circ}\xspace}

\newcommand{\eat}[1]{}

\newcommand{\cis}{{C\&IS}\xspace}

\begin{document}

\setlength{\textfloatsep}{9pt}

\title{
Efficient Detection of Points of Interest from Georeferenced Visual Content
}

\author{
Ying Lu\inst{1}\thanks{Ying Lu did most of this work as an intern at
Samsung Research America.} 
\and 
Juan A. Colmenares\inst{2}
}
\authorrunning{Y. Lu and J.A. Colmenares} 
%
\tocauthor{Ying Lu, Juan A. Colmenares}
\institute{
University of Southern California, USA\\
\email{ylu720@usc.edu} 
\and
Samsung Research America, USA\\
\email{juan.col@samsung.com} 
}

\maketitle

\begin{abstract}
Many people take photos and videos with smartphones and more recently with
360\degree cameras at popular places and events, and share them in social media. 
Such visual content is produced in large volumes in urban areas, and it is a
source of information that online users could exploit to learn what has got the
interest of the general public on the streets of the cities where they live or
plan to visit.
A key step to providing users with that information is to identify the most
popular k spots in specified areas.
In this paper, we propose a clustering and incremental sampling (\cis) approach
that trades off accuracy of top-k results for detection speed.
It uses \emph{clustering} to determine areas with high density of visual content, and 
\emph{incremental sampling}, controlled by stopping criteria,  
to limit the amount of computational work.
It leverages spatial metadata, which represent the scenes in the visual content,
to rapidly detect the hotspots, and uses a recently proposed Gaussian
probability model to describe the capture intention distribution in the query
area. 
We evaluate the approach with metadata, derived from a non-synthetic,
user-generated dataset, for regular mobile and 360\degree~visual content.
Our results show that the \cis approach offers 2.8$\times$--19$\times$
reductions in processing time over an optimized baseline, while in most cases
correctly identifying 4 out of 5 top locations.  
\end{abstract}

\section{Introduction} \label{sec:intro}


Many people frequently use their smartphones and more recently their
360$^{\circ}$ cameras to take photos and videos to capture memorable subjects
and situations at places and events (\eg touristic attractions, concerts, and
political rallies). 
People also upload their visual content (\ie photos and videos) very often to
social media websites, such as Facebook, Flickr, Instagram, and YouTube.
Urban areas naturally produce visual content in large volumes. 
A recent study~\cite{nyc-mobile-study15} indicates that over 75\% of 
people in New York City (NYC) own smartphones;  
\ie $\sim$6.3 million people 
and $\sim$0.3\% of roughly 2 billion smartphone users worldwide~\cite{kissonergis15}.
With 350+ million photos and videos being uploaded daily to
Facebook~\cite{lowe16}, NYC may produce over 1 million pieces of visual content
per day.

Such abundant and continuously generated visual content offers online users the
opportunity to learn about subjects, places, and events that have caught the
attention of people (physically present) in a given area. 
For example, users often search the web to know 
\emph{what popular attractions are currently in their city} 
as well as   
\emph{what interesting events have recently happened there}.
Most web services today answer this type of questions by looking at camera
locations and timestamps photos and videos have been tagged with.
The results, however, are inherently imprecise because the camera and the
subject are usually at different locations and many times far apart (\eg
pictures of the Statue of Liberty are usually taken at a considerable distance
from it). 
In addition, travel and review websites like TripAdvisor and Yelp are often
limited to well-known, static landmarks, but interesting events can also happen
at ad hoc locations (\eg an amazing musical performance down the street).

In this paper, we focus on efficiently identifying the top-k most popular
\emph{points of interest} (POIs) from photos and videos.
POIs are estimated locations of subjects captured in the visual content; the
subjects' locations are obtained from analyzing (metadata of) visible scenes.
We take into account that POIs are not necessarily static and their
appeal may vary over time (\eg the Barra Olympic Park, in Rio de Janeiro, was
the world's focus during the 2016 Summer Olympics, but was abandoned after six
months~\cite{rio-olympic-news}); so, we allow POIs to be identified in specific time
intervals.

\smallskip
\noindent\textbf{Background and Baseline Approach}~
Early approaches in the literature (\eg~\cite{duygulu-eccv2002,sivic-iccv2005})
identify POIs from visual content by extracting and analyzing image features. 
They are computationally intensive and thus not applicable to large volumes of
visual content. 
To accelerate the process, researchers have proposed other approaches 
(\eg~\cite{zheng-cvpr2009,ji-mm2009,kennedy-mm2007,liu-neurocomputing2013,hao-mm2011,hao-tmm2014,zhang-tmm2016}) 
that leverage sensor-generated geo-metadata (\eg GPS locations, timestamps, and
compass directions) associated with the visual content. 
From this group, most of the studies (\eg~\cite{zheng-cvpr2009,ji-mm2009,
kennedy-mm2007,liu-neurocomputing2013}) detect hotspots based on camera
locations. 
However, using the camera location is insufficient to represent the coverage
of a photo or video. 
On one hand, as mentioned above, the location of the camera and the location of
the subject in a photo or video are often not the same and many times are far
apart~\cite{hao-mm2011,hao-tmm2014,zhang-tmm2016}. 
On the other hand, cameramen often move during video recording; thus, a single
camera location is inadequate for a video with a trajectory.
To avoid this issue, recent studies on POI
identification~\cite{hao-mm2011,hao-tmm2014,zhang-tmm2016} represent the visible
scene of a photo or individual video frames with the spatial extent of its
coverage area at a fine granularity (\ie geo-tagged at the video frame level). 
Such spatial extent of a scene is called \emph{field of view}
(\fov)~\cite{ay-mm2008} and is illustrated in \fref{fig:fov}.

Among the studies based on the \fov model, the most recent
approach~\cite{zhang-tmm2016}
-- the state-of-the-art method -- 
identifies POIs from georeferenced videos by partitioning the query area into 
a grid with equally-spaced cells and using a Gaussian probability model to 
describe the capture intention distribution on the grid. 
This approach, described in \sref{sec:baseline}, has been shown to be able to
achieve high accuracy ($\leq$1~m), and for that reason we adopt it as our
baseline. 
However, the baseline approach, if implemented naively, takes long time to
process big areas with large volumes of visual content as it computes the
capture intention contribution of \emph{all} \fovs to \emph{every} cell in the
query area (\eg over 1 hour to detect the top-5 spots in Munich and 19 hours
in Los Angeles).
Such long processing time would render the approach incapable to offer an
interactive user experience.

\smallskip
\noindent\textbf{Challenge}~ 
The challenge in this work is how to accelerate top-k POI detection without
much loss of \emph{accuracy} when compared to the results of the baseline approach.
In other words, how to trade off detection efficiency and accuracy.  
One may attempt to reduce the number of cells to be processed by increasing the
cell size. 
This simple method can reduce the detection time proportionally to the
cell count; however, it may significantly deteriorate the result accuracy mainly
because larger cells cover the target area with a lower resolution (see details
in \sref{sec:base-alg-problems}).
Another option is to reduce the number of \fovs simply by using random sampling.
But, determining the right sample size that offers good detection speed and
accurate results across multiple target areas is not straightforward. 

\smallskip
\noindent\textbf{Contributions}~ 
To overcome the challenge, we propose a series of techniques by exploiting the 
spatial properties of \fovs.
Our contributions are as follows:
\begin{compactenum}
\item
We first introduce two practical optimization techniques
(\sref{sec:optimizations}) that enable significant (up to 2000$\times$)
reduction in detection time with no accuracy loss.
The first technique seeks to reduce the number of grid cells processed per \fov;
thus, it only considers the cells that overlap the \emph{minimum bounding
rectangle} (MBR) of each \fov, rather than the entire grid. 
The second technique makes an adjustment to the probability model to properly
handle and efficiently process 360\degree~visual content, which is proliferating
as 360\degree~cameras and virtual reality headsets have entered the market.
Both techniques are combined in an improved implementation of the adopted
baseline approach, referred to as the \emph{optimized baseline}.

\item
Considering that densely populated areas are expected to contain 
very large number of \fovs and a fraction of the \fovs may be enough to 
identify the top-k spots, we propose an approach that combines two well-known
techniques:  
1)~\emph{clustering} to determine areas with high density of \fovs, where the
hotspots are more likely to be, 
and 
2)~\emph{incremental sampling} to limit the number of \fovs to be processed and
thus reduce detection time.
This clustering and incremental sampling (\cis) approach, described in
\sref{subsec:iter-sampling}, relies on stopping criteria to make incremental
sampling terminate after having some indications that the top-k results have
been identified.
By combining these techniques, we can flexibly trade off result accuracy and
detection speed.

\item
We conducted extensive experiments with 
a real-world geo-tagged video dataset~\cite{lu-mmsys2016}
recorded with regular mobile phones
plus two variants of the same dataset modified assuming the use of both 
regular and 360$^{\circ}$ cameras. 
Experimental results demonstrate that 
the \cis approach brings 2.8$\times$--19$\times$ improvements in
processing time over the optimized baseline, while in most cases correctly
identifying 4 out of 5 top locations.  

\end{compactenum}

\section{Field of View} \label{sec:fov}

\begin{figure}[!tbp]
\centering
\includegraphics[width=0.20\textwidth]{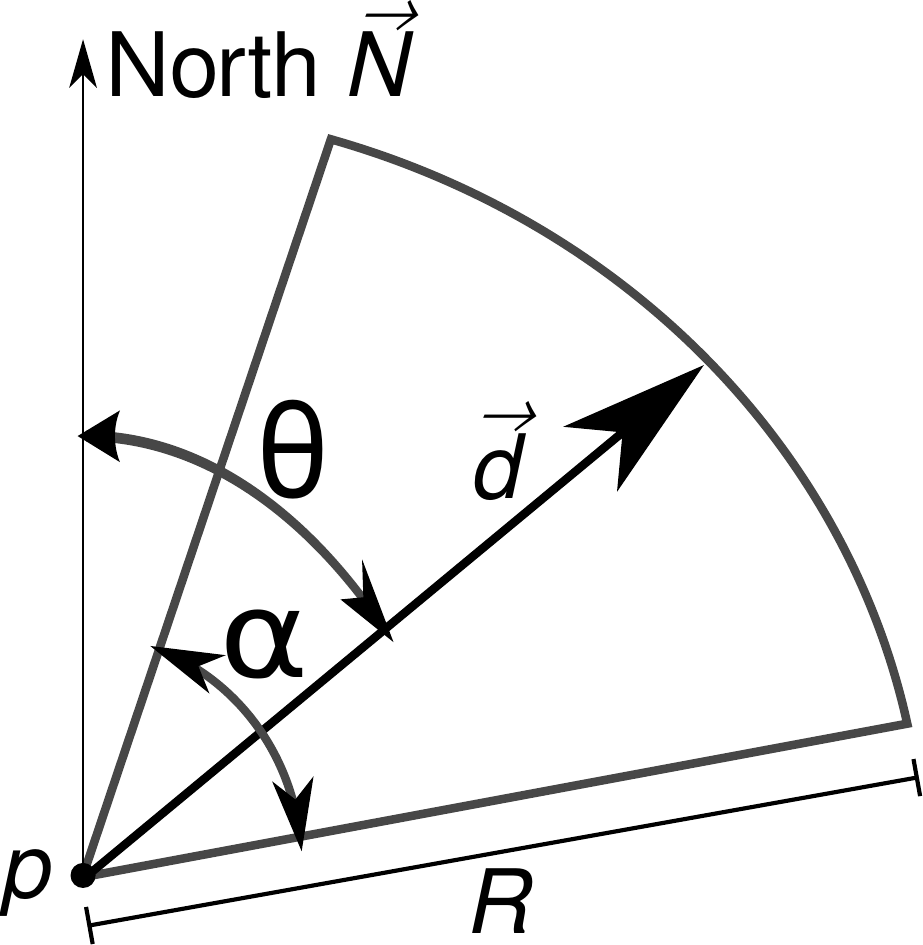}
\vspace*{-0.1in}
\caption{Two-dimensional field of view (2D \fov).}
\label{fig:fov}
\end{figure}


Visual content can be captured along with metadata representing the scenes in
it, particularly their spatial features. 
This is easily achievable today with smartphones equipped with cameras and all
the necessary sensors.
A photo and individual frames in a video (\eg a frame every second in a 30-fps
video) can be tagged with a \emph{field of view} (\fov)~\cite{ay-mm2008}, a
piece of metadata that describes the area covered by the captured scene.
A two-dimensional \fov, illustrated in \fref{fig:fov}, is stored as a tuple
$\langle p,\theta,R,\alpha \rangle$, where 
$p$ is the camera's location given by its latitude and longitude coordinates,
$\theta$ is the camera's orientation or azimuth (\ie 
the angle between the north reference line and the camera's shooting direction
$\vec{d}$ measured clockwise), 
$R$ is the maximum visible distance from $p$ at which an object within the
camera's \fov can be recognized,
and 
$\alpha$ is the camera's visible angle.
Moreover, \fovs may contain an extra field $t$ with the time at which the scene
was captured.

With a smartphone or another sensor-rich camera device, 
$p$ and
$\theta$ can be read from the GPS receiver and digital compass,
respectively, and 
$\alpha$ can be obtained based on the properties of the camera and its lens
for the current zoom level~\cite{graham65}. 
In addition, 
the maximum visible distance can be estimated with the formula 
$R = f.h/y$, where 
$f$ is the camera's focal length, 
$h$ is the height of the visible object with maximum depth of view, 
and 
$y$ is the object's height in the image, which is typically at least
$1/20^{th}$ of the image's height~\cite{ay-mm2008}.

Two-dimensional (2D) \fovs are pie-shaped ($\alpha < 360\degree$) or circular
($\alpha = 360\degree$); they assume that the camera and target are on the same
plane, and only consider azimuth rotation (\emph{yaw} movements).
By contrast, \fovs in a three-dimensional space are cone-shaped ($\alpha <
360\degree$) or spherical ($\alpha = 360\degree$), and consider, besides yaw,
the other two rotation axis: \emph{pitch} and \emph{roll}. 
In this work, we focus on 2D \fovs.

\section{Baseline Approach} 
\label{sec:baseline}

An approach to detecting points of interest from georeferenced videos has been
recently proposed in~\cite{zhang-tmm2016}.
It uses 2D \fovs to represent visible scenes and applies a Gaussian probability
model to describe the capture intention distribution in a user-specified area. 
It has been shown to achieve high accuracy (\ie a meter or less between detected
hotspots and their actual locations). 
We adopt this state-of-the-art approach as our \emph{baseline}.

The approach first partitions the user-specified area $A$ into a grid with
equally-spaced cells. 
It assumes the centers of the cells are the visual targets (\ie each cell is
represented by its center). 
Then, it calculates the capture intention of each cell $c$ as follows:
\begin{eqnarray}
\gamma(c,F) = \frac{\sum_{f \in F}ci(c,f)}{max_{c \in C}\sum_{f \in F}ci(c,f)}
\label{eqn:cell-allover-ci}
\end{eqnarray}
$\gamma(c,F)$ is the normalized sum of the individual intentions of the
\fovs covering the area $A$ to capture the cell $c$. 
The set of \fovs overlapping $A$ is denoted as $F$, and the set of cells forming
the grid as $C$. 
Finally, the $k$ cells with the highest cumulative capture intention are
returned as the top-$k$ points of interest. 

\begin{figure}[!tbp]
\centering
\includegraphics[width=0.20\columnwidth]{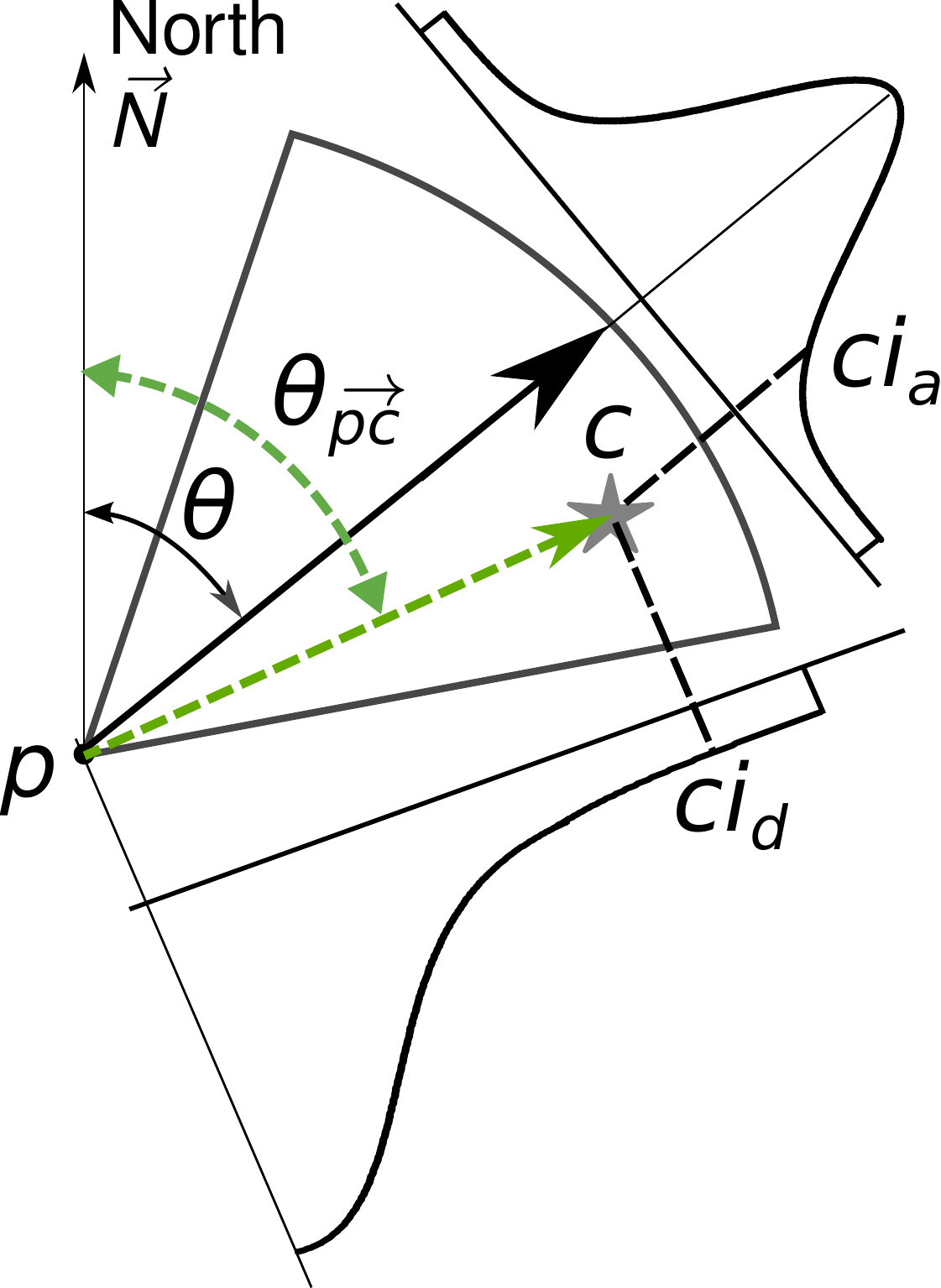}
\vspace*{-0.1in}
\caption{
An \fov{'s} contribution to the capture intention of a cell $c$ involves 
i)~$ci_a$, the capture intention probability with respect to the angular
difference $|\theta - \theta_{\vec{pc}}|$, 
and 
ii)~$ci_p$, the capture intention probability with respect to the distance
$\|p,c\|$. 
}
\label{fig:ci-factors}
\end{figure}

The individual contribution of an \fov $f$ to the capture intention of a cell
$c$ involves two factors (see \fref{fig:ci-factors}):
the \emph{angular difference} $|\theta - \theta_{\vec{pc}}|$ between 
the camera's shooting direction $\vec{d}$ and the cell's center, 
and
the Euclidean \emph{distance} $\|p,c\|$ between the camera's location and the cell's center. 
Such contribution is calculated as: 
\begin{eqnarray}
ci(c,f) = ci_a(c,f) \times ci_d(c,f)
\label{eqn:cell-onefov-ci}
\end{eqnarray}
where $ci_a$ is the probability of capture intention with respect to the angular
difference,
and 
$ci_d$ is the probability of capture intention with respect to the distance.

In general, people are more likely to capture the target in the center of the
image (along $\vec{d}$). 
Intuitively the capture intention increases as the angular difference
decreases.  
For that reason, the approach adopts a Gaussian distribution to model 
the angular capture intention $ci_a(c,f,\sigma_a)$ that an \fov $f \triangleq \langle
p,\theta,R,\alpha \rangle$ has for a cell $c$. 
The distribution is given by:
\begin{eqnarray}
ci_{a}(c,f,\sigma_a) = 
\begin{cases}
\frac{e^{-\frac{(\theta-\theta_{\vec{pc}})^2}{2\sigma_a^2}}}{\sqrt{2\pi}\sigma_a}: & 
    |\theta-\theta_{\vec{pc}}|\geq \frac{\alpha}{2} \\
0: &  \text{otherwise}
\end{cases},
\label{eqn:ci-angle}
\end{eqnarray}
where $\sigma_a$ is the variance.

Similarly, people tend to capture the target close to the camera for better
visibility; so, the closer the target the higher the capture intention. 
In this case, the approach models the distance-based capture intention
$ci_{d}(c,f,\sigma_d)$ of an \fov $f$ on a cell $c$ with the following
distribution: 
\begin{eqnarray}
ci_{d}(c,f,\sigma_d) = 
\begin{cases}
\frac{e^{-\frac{\|p, c\|^2}{2\sigma_d^2}}}{\sqrt{2\pi}\sigma_d}: &\|p, c\| \leq R\\
0: &  \text{otherwise}
\end{cases},
\label{eqn:ci-dist} 
\end{eqnarray}
where $\sigma_d$ is the variance.

%
Algorithm~\ref{alg:baseline} presents the pseudocode of the baseline 
approach~\cite{zhang-tmm2016}. 
The \textsc{calcCIMatrix} procedure computes the capture intention matrix of
\fovs. 
Specifically, line~\ref{calc_capture_intention} calculates the individual
contribution of an \fov to the capture intention of a cell, using the Equations
(\ref{eqn:cell-onefov-ci}), (\ref{eqn:ci-angle}) and (\ref{eqn:ci-dist}).

%

\begin{figure}[!tbp]
\centering
\begin{minipage}[t]{0.30\linewidth}
\subfigure [Full-coverage~\cite{hao-tmm2014}] {
\includegraphics[width=0.55\columnwidth]{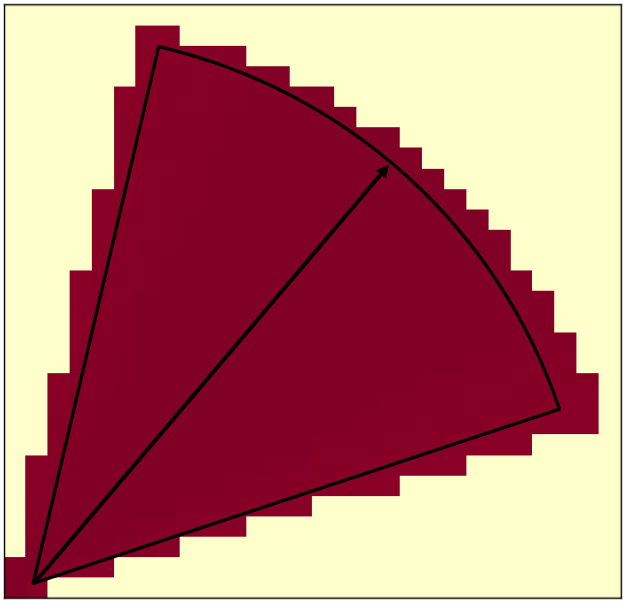}
\label{fig:full-coverage-model}
}
\end{minipage}\hspace{1mm}
\begin{minipage}[t]{0.30\linewidth}
\subfigure [Center-line~\cite{hao-tmm2014}] {
\includegraphics[width=0.55\columnwidth]{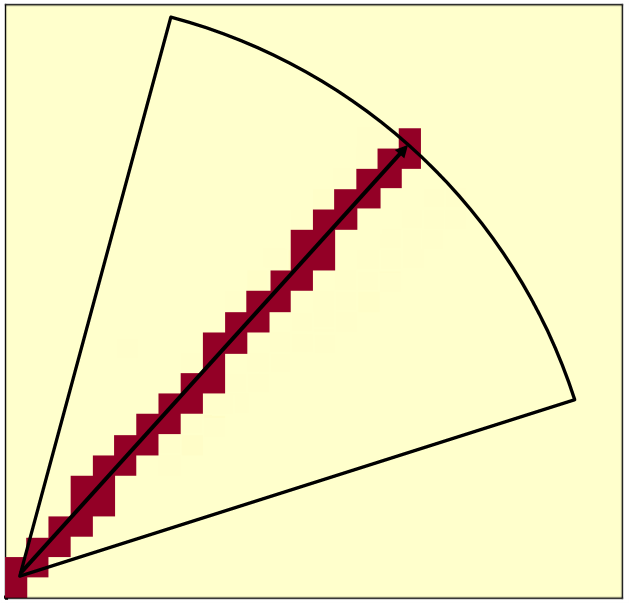}
\label{fig:center-line-model}
}
\end{minipage}\hspace{1mm}
\begin{minipage}[t]{0.30\linewidth}
\subfigure [Probabilistic~\cite{zhang-tmm2016}] {
\includegraphics[width=0.55\columnwidth]{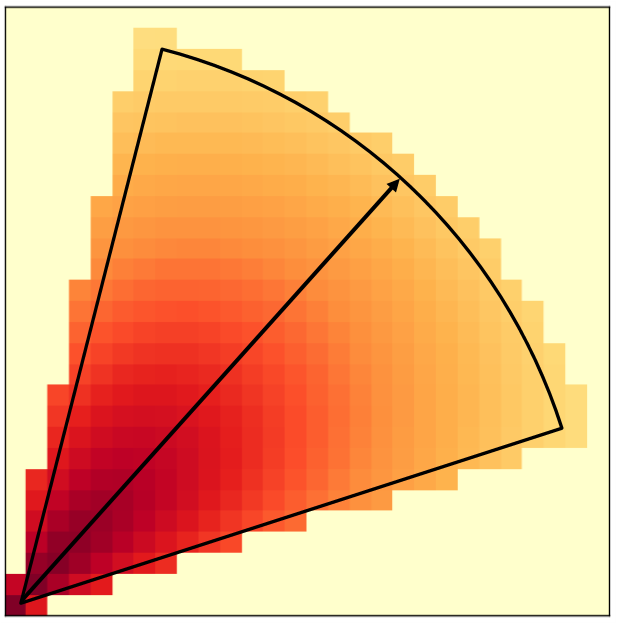}
\label{fig:probability-model}
}
\end{minipage}
\vspace{-2mm}
\caption{
Models for estimating the capture intention contributions of an \fov.
The darker the color of a cell, the higher its capture intention value. 
}
\label{fig:different-ci-models}
\end{figure}


This Gaussian probability model is the key distinguishing feature
of the state-of-the-art approach we have adopted as our
baseline~\cite{zhang-tmm2016}.
\fref{fig:different-ci-models} illustrates the existing models for calculating
the capture intention contributions of an \fov.
The full-coverage model~\cite{hao-mm2011,hao-tmm2014} gives the same value to
every cell (or point) inside an \fov.
The center-line model~\cite{hao-mm2011,hao-tmm2014} only considers the cells
in the \fov along the shooting direction.
With the probabilistic model, as shown in Figure~\ref{fig:probability-model}, 
capture intention values smoothly spread over the space inside the \fov.
In particular, the contribution from an \fov to the capture intention of a cell
increases as the cell gets closer to the camera's location and to the camera's
shooting direction. 
Research indicates that people tend to focus on the center of an
image~\cite{judd-iccv2009}. 
Additionally, a closer object is likely to be more prominent in an image or
video frame.
Hence, the probabilistic model can yield better accuracy than the other two
models. 
For example, at Merlion Park (Singapore), with the probabilistic model, 
the distances between the detected POIs and their ground-truth locations are less 
than 0.8 meters, while the distance error with the center-line model, which 
outperforms the full-coverage model~\cite{hao-tmm2014}, is more than 
35 meters~\cite{zhang-tmm2016}.

\begin{algorithm}[!tbp]
\begin{algorithmic}[1]
\scriptsize

\Require 
\Statex $k$: Number of top cells to detect.
\Statex $A$: Area of interest, specified by its  
minimum and maximum latitude, and minimum and maximum longitude.
\Statex $T$: Time interval of interest. 
\Statex $l$: Length of each side of the cell forming the grid.  
\Statex $\sigma_{a}$: Variance of angular capture intention distribution.
\Statex $\sigma_{d}$: Variance of distance-based capture intention distribution. 

\Ensure 
\Statex $K$: Set with the top-k cells. 
\Statex

\State $F \gets$ \Call{getFoVsInRange}{$A$, $T$} 
\Comment{Range query to obtain set of \fovs.}

\State Determine the size of a matrix $M$: $xdim \times ydim$ from $A$ and $l$
\State $M \gets$ \Call{ZeroMatrix}{$xdim$, $ydim$} 
\Comment{Initialize matrix.} 

\State \Call{calcCIMatrix}{$F$, $\sigma_{a}$, $\sigma_{d}$, 0, $xdim$, 0, $ydim$, $M$} 
\Comment{Calculate the caption intention matrix.}


\State $K \gets$ \Call{getTopKCells}{$M$} 
\Comment{Obtain top-k cells with the highest capture intentions from the matrix.} 
\State return $K$ 

\Statex
\Procedure{calcCIMatrix}{$F$, $\sigma_{a}$, $\sigma_{d}$, 
                         $x_{min}$, $x_{max}$, $y_{min}$, $y_{max}$, $M$}
\For{each $f$ in $F$}
  \For{$y \gets y_{min}$ to $y_{max}$}
    \For{$x \gets x_{min}$ to $x_{max}$}
      \State Let $center$ be the center of the cell ($x$, $y$)
      \State $ci \gets$ \Call{calcCaptureIntention}{$f$, $center$, $\sigma_{a}$, $\sigma_{d}$}
      \label{calc_capture_intention}
      \If{$ci > 0$}
        \State $M(x,y) \gets M(x,y) + ci$
      \EndIf
    \EndFor
  \EndFor
\EndFor
\EndProcedure
\end{algorithmic}
\caption{
\small 
Naive baseline approach.
}
\label{alg:baseline}
\end{algorithm}

\subsection{Drawbacks}
\label{sec:base-alg-problems}

Despite the high accuracy, the naive baseline has a
major drawback: 
The algorithm (\ie the procedure \textsc{calcCIMatrix}) takes long time to process big areas with
large number of \fovs since it computes the capture intention contribution of
\emph{all} \fovs to \emph{every} cell in the query area.
Its time complexity is $\mathcal{O}(N_c \times N_f)$, where
$N_c$ is the number of cells in the grid, and 
$N_f$ is the number of \fovs.
For example, according to our evaluation (see \sref{sec:eval}), 
it takes over 19 hours to find the top-5 cells in Los Angeles area
($\sim$1,300~km$^2$) with over 12.6 million cells of $\sim$11m$\times$11m and
about 52,000 \fovs from the user-generated dataset GeoUGV~\cite{lu-mmsys2016}.
Back in 2009, it was estimated that 4,306 photos were uploaded daily to Flickr
from Los Angeles~\cite{crandall-www2009}.  
Assuming proportionality to \fov count, the detection time of the top-5 cells in
that case would be over 1.5 hours; 
\ie the user may need to wait way more than an hour for a visual summary of what
happened in Los Angeles in the last 24 hours.
Such long processing time is clearly unacceptable for interactive user
experience.

\begin{figure}[!tbp]
\centering
\subfigure [Cell size: 55m$\times$55m] {
\includegraphics[width=0.25\columnwidth]{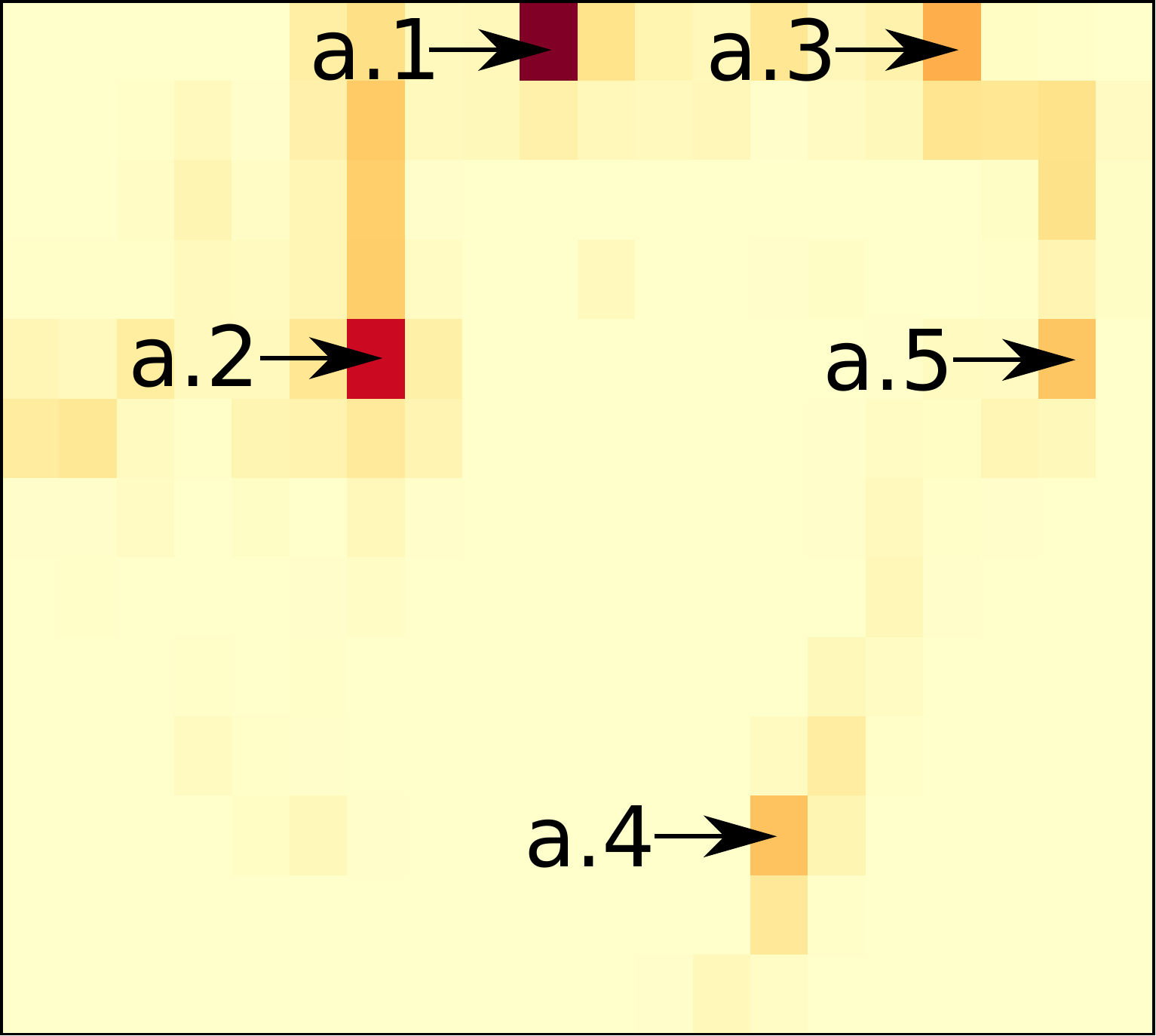}
\label{fig:cellsize-55}
}
\hspace{10ex}
\subfigure [Cell size: 111m$\times$111m] {
\includegraphics[width=0.25\columnwidth]{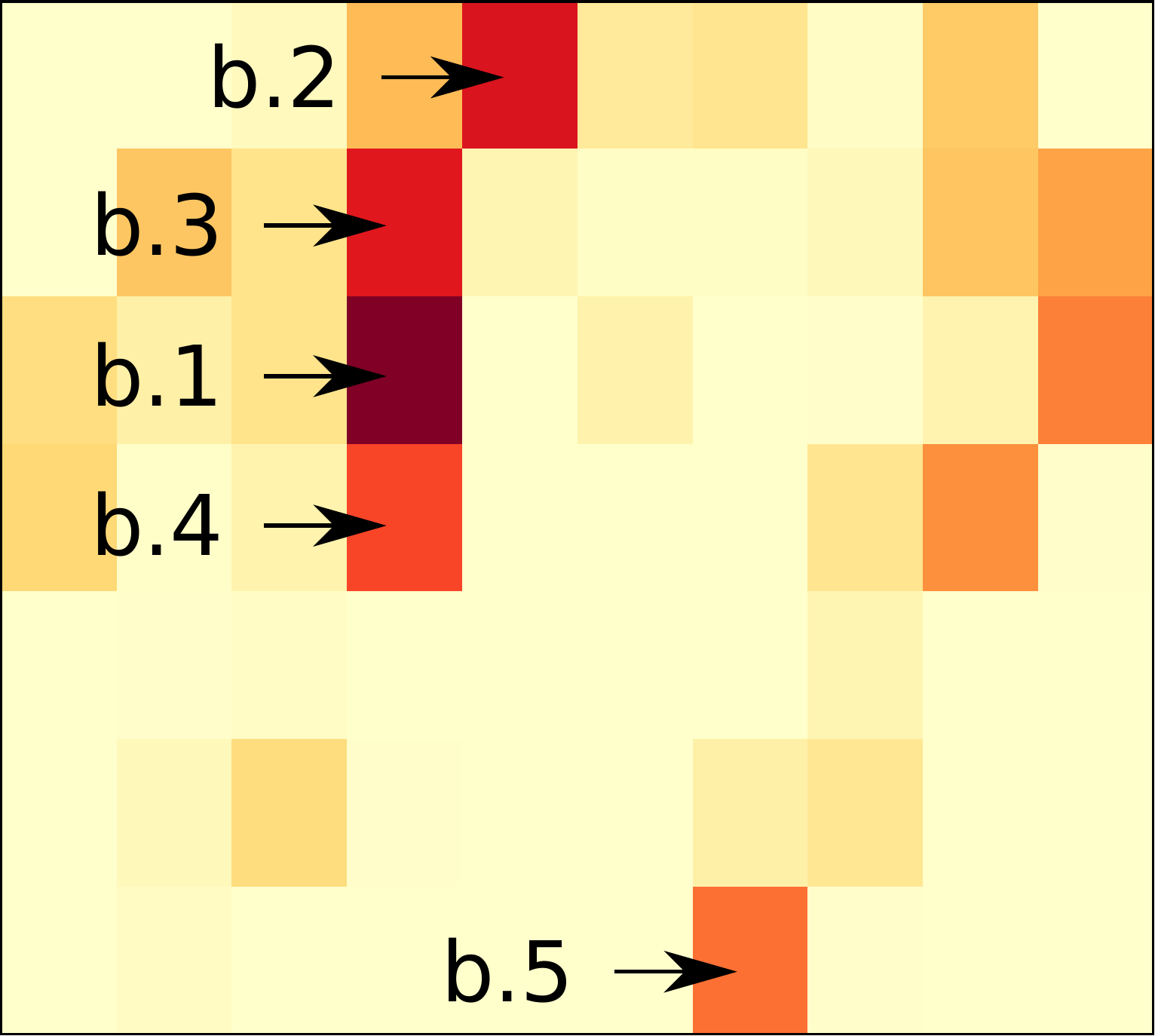}
\label{fig:cellsize-111}
}
\vspace*{-0.1in}
\caption{
Different top-5 results with two cells sizes at Merlion Park.
}
\label{fig:effect-of-cell-size}
\end{figure}

\begin{figure}[!tbp]
\centering
\includegraphics[width=0.60\columnwidth]{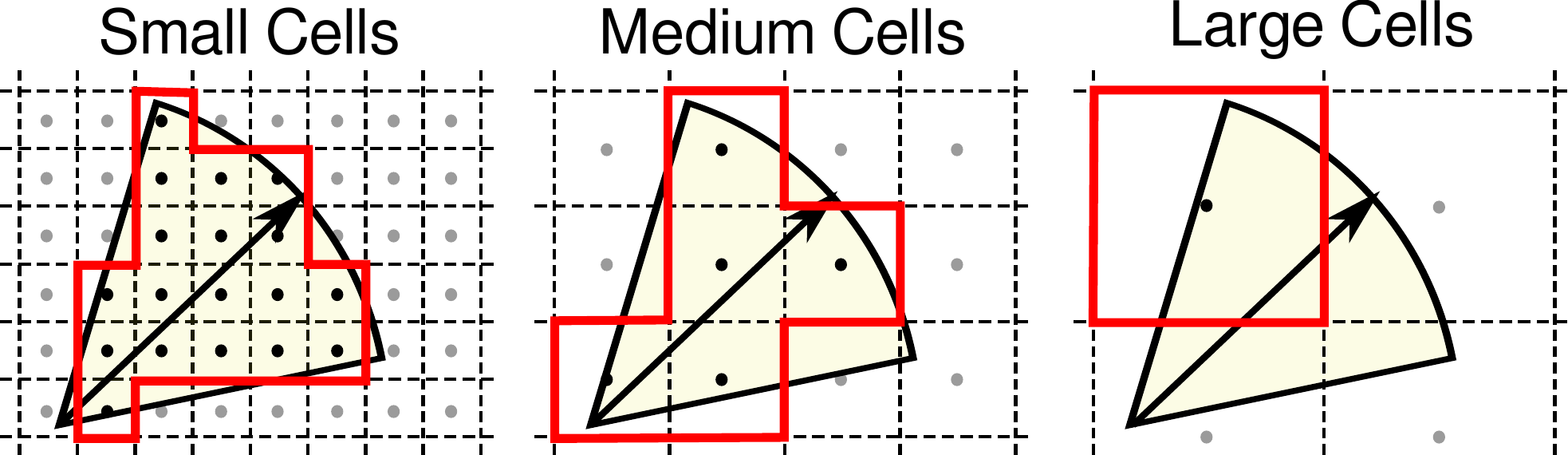}
\vspace*{-0.1in}
\caption{
As the cell sizes increase, the centers of the cells covered by an \fov change,
along with the polygon those cells form. 
}
\label{fig:fov-misses-cell-centers}
\end{figure}

Since the approach's complexity is $\mathcal{O}(N_c \times N_f)$, its detection
time can be shortened by reducing the number of cells ($N_c$) and/or the number
of \fovs ($N_f$) that need processing. 
An easy way to reduce $N_c$ is to increase the cell size.
Unfortunately, while it reduces the detection time proportionally to the cell
count, the accuracy may deteriorate significantly. 
For example, \fref{fig:effect-of-cell-size} shows the capture intention heatmaps
at Merlion Park (Singapore) for two cell sizes. 
Looking at the discrepancies between the two top-5 result sets,
\fref{fig:cellsize-111}, with larger cells, 
misses the 3rd and 5th cells ($a.3$ and $a.5$) in \fref{fig:cellsize-55} and 
misidentifies its 3rd and 4th cells ($b.3$ and $b.4$).
The reason is that the approach works best on a fine grid, with cells much
smaller than the \fovs. 
This way cells can record high-resolution changes in capture intention on the
grid.
As the cell size increases, the centers of the cells covered by an \fov change,
along with the polygon those cells form  (see
\fref{fig:fov-misses-cell-centers}).
And, since a cell's capture intention depends on the location of its center, the
capture intention of a large cell is in general not equal to the sum of the
capture intentions of smaller adjacent cells that make up the large cell.
In addition, in practice it is difficult to set a single cell size that properly
balances processing time and accuracy because \fovs come in different sizes (\ie
their parameters $R$ and $\alpha$ vary).

\section{Optimized Baseline} \label{sec:optimizations}

In this section, we propose two practical techniques to accelerate
the baseline approach with no need to increase the cell size. 
The optimized procedure implementing these techniques is denoted as
\textsc{calcCIMatrix$^{+}$}, and it is interchangeable with the naive procedure
in Algorithm~\ref{alg:baseline} as they both receive the same parameters.


\begin{figure}[b]
\centering
\includegraphics[width=0.35\columnwidth]{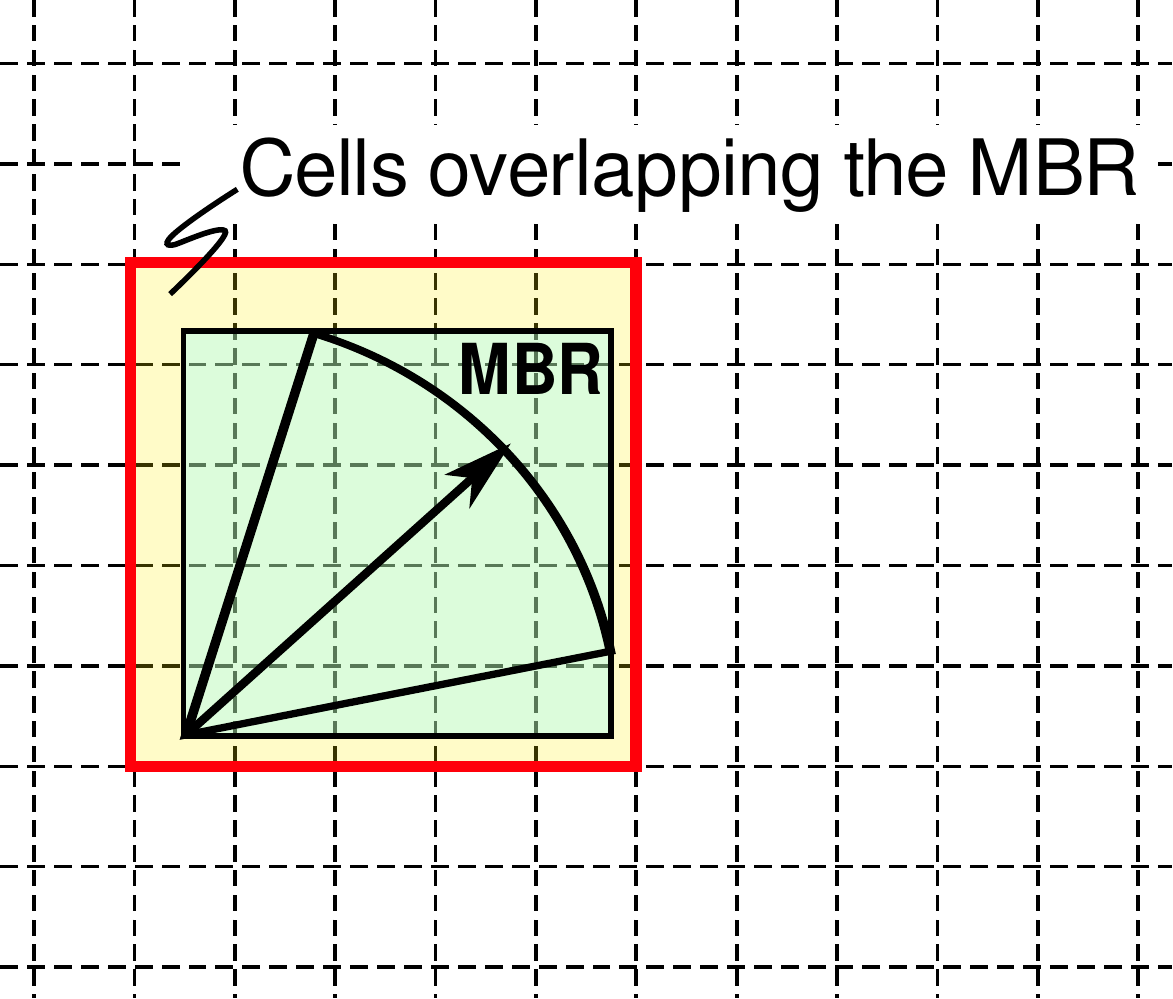}
\vspace*{-0.1in}
\caption{
MBR-based cell filtering. 
For each \fov, our approach only considers the cells that overlap the upright minimum
bounding rectangle (MBR) of the \fov.
}
\label{fig:cells-overlapping-mbr}
\vspace*{-0.1in}
\end{figure}

\subsubsection{MBR-based Cell Filtering} \label{sec:mbr-based-cell-filtering}
This technique seeks to reduce the number of cells ($N_c$) to be
processed per \fov.
The idea is to only compute the capture intention probability of the cells that
fully or partially overlap the \emph{upright minimum bounding rectangle} (MBR)
of each \fov (see \fref{fig:cells-overlapping-mbr}).
The reason is that an \fov surely makes no contribution to the capture intention
of the cells not covered by its MBR; those cells can then be ignored 
when processing the \fov.

This simple, yet effective optimization exploits the fact that each \fov often
covers a small fraction of the whole area of interest. 
For example, Los Angeles and Munich metropolitan areas cover 12.5E9~m$^2$ and
27.7E9~m$^2$, respectively; by contrast, a circular \fov with a
maximum visible distance of 100~m only covers 31.4E3~m$^2$ (six orders of
magnitude less).
%
%
%
This technique requires \fovs to be augmented with their upright MBRs.
MBRs are aligned with the geographic coordinate system and comprise four
coordinate values each (maximum and minimum latitude, and maximum and minimum
longitude). 
One option is to store the \fovs along with their MBRs, where MBRs are either
provided by the camera devices or calculated by the storage system at ingestion
time.
Another option is to calculate the \fovs' MBRs on the fly at query time. 
We consider that storing \fovs with their MBRs is more beneficial given the
significant speedups this optimization brings.
Our evaluation (\sref{sec:eval-mbr-cell-filt}) shows that MBR-based cell
filtering reduces the processing time up to three orders of magnitude when
compared to the naive baseline. 


\subsubsection{Improving Efficiency for 360$^{\circ}$ Visual Content}
\label{sec:impr-4-360-content}
The baseline approach can be easily improved even further to process
360$^{\circ}$ visual content more efficiently.
The main observation is that the intention of a circular \fov (with visible
angle $\alpha = 360^{\circ}$) is to capture its entire surrounding. 
That is, the angular capture intention of a circular \fov is constant and
equal to 1 in every direction ($ci_{a} \triangleq 1$).
Hence, $ci_{a}$ need not be computed for the cells covered by circular \fovs,
saving some time in detecting the top-k cells.

\section{Clustering and Incremental Sampling (\cis)} 
\label{subsec:iter-sampling}

MBR-based cell filtering reduces the detection time of the baseline approach by
limiting the number of cells that need to be processed per \fov.
But densely populated areas are expected to contain very large number of \fovs,
and a fraction of the \fovs may be sufficient to identify the 5 to 10 top spots.
A simple approach to limit the number of \fovs is to first take a sample,
uniformly distributed and without repetitions, of the \fov population in the
query area, and then call \textsc{calcCIMatrix$^{+}$} to identify the top-k
cells using the sample.
We call this approach \emph{single sampling}. \label{sec:single-sampling}
Clearly, the less \fovs in the sample, the faster the results are produced, 
but the less accurate those results are likely to be. 
Single sampling is not flexible in trading off efficiency and accuracy for two
reasons. 
First, it samples the \fov population once with a predetermined sample size.
It is difficult to determine the optimal sample size to achieve the best
performance (\ie fast detection time and high accuracy) for a given area. 
Second, the single sampling approach samples all the \fovs in the entire target
area. 
Different datasets and areas may have different density distributions of
\fovs; thus, their optimal sample sizes may be very different. 
To overcome these drawbacks, we proceed to present our 
approach that is based on clustering and incremental sampling techniques.

\begin{algorithm}[!tp]
\begin{algorithmic}[1]

\smallskip

\Require 
\Statex $V$: Video database. 
\Statex $k$: Number of top cells to detect.
\Statex $A$: Area of interest. 
\Statex $T$: Time interval of interest. 
\Statex $l$: Length of each side of the cell forming the grid.  
\Statex $\sigma_{a}$: Parameter for Gaussian distribution of angular capture intention.  
\Statex $\sigma_{d}$: Parameter for Gaussian distribution of distance capture intention.  
\Statex $c$: Number of clusters to identify. 
\Statex $f_c$: Fraction of the \fov population used in cluster identification.
\Statex $f_i$: Fraction of the \fov population used in each iteration.

\smallskip

\Ensure 
\Statex $K$: Set with the top-k cells. 
\Statex

\smallskip

\State $F \gets$ \Call{getFoVsInRange}{$V$, $A$, $T$} \label{cis:range-query-all}
\Comment{Query to obtain the set of \fov{s}.}



\State $S \gets$ \Call{getRandomSampleNoRepetitions}{$F$, $f_c$} \label{cis:sampling}

\State $C \gets$ \Call{identifyClusters}{$S$, $c$}  \label{cis:clustering}

\State $K \gets$ \Call{Heap}{$k$} 
\Comment{Initialize the global heap for top-k cells.} 

\For{each $cluster$ in $C$} \label{cis:BEGIN-for-each-cluster}

  \State $radius \gets$ \Call{calcRMSRadius}{$cluster$}
  \Comment{Root Mean Square (RMS) radius}

  \State $A_c \gets$ \Call{getBoundingRect}{$cluster.center$, $radius$}
  
  \State [$x_{min}$,$x_{max}$,$y_{min}$,$y_{max}$] $\gets$ \Call{calcCellRange}{$A$, $A_c$, $l$}

  \State $F_c \gets$ \Call{getFoVsInRange}{$F$, $A_c$, $T$} 
  \label{cis:range-query-cluster}
 
  \State $M_c \gets$ \Call{ZeroMatrix}{$xdim$,$ydim$}
  
  \State $iter \gets 0$

  \Do \label{cis:BEGIN-iter-sampling}

    \State $S' \gets$ \Call{getRandomSampleNoRepetitions}{$F_c$, $f_i$} 
    \label{cis:iter-sampling} 

    \State $M' \gets$ \Call{ZeroMatrix}{$xdim$,$ydim$}

    \State \Call{calcCIMatrix$^{+}$}{$S'$, $\sigma_{a}$, $\sigma_{d}$, 
                 $x_{min}$, $x_{max}$, $y_{min}$, $y_{max}$, $M'$} \label{cis:detect}

    \State $M_c \gets M_c + M'$
    \Comment{Update matrix for the cluster.} 
 
    \State $K_c \gets$ \Call{getTopKCells}{$M_c$}
    \Comment{Obtain top-k cells for the cluster.} 

    \State Exclude $S'$ from $F_c$ 
    \Comment{Same \fovs will not be reused at different iterations.}

    \State $iter \gets iter +1$    

  \doWhile{\Call{satisfyStopCriteria}{$K_c$, $M_c$, $iter$}} 
  \label{cis:stopping-criteria}\label{cis:END-iter-sampling}

  \State $p \gets min(iter \times f_{i}, 1)$ 
  \Comment{The percentage of \fovs in the cluster that are considered.}
  \label{cis:pop-fraction}


  \State \Call{$K$.update}{$K_c$, $p$} \label{cis:update-global-heap}
  \Comment{Update the global top-k cells.} 

\EndFor \label{cis:END-for-each-cluster}



\State return $K$

\end{algorithmic}
\caption{Clustering and incremental sampling approach.}
\label{alg:clustering-and-iter-sampling}
\end{algorithm}

%
People usually take photos and videos at the same locations (\eg
touristic attractions, stadiums, and concert venues). 
So there is no surprise that social media visual content is often heavily
concentrated around specific spots.
We leverage this by focusing on areas with high density of \fovs, where the
top-k cells are more likely to be, and ignoring sparse areas. 
This, in turn, helps reduce the number of cells to be processed and, to a lesser
extent, the number of \fovs. 

We apply a \emph{clustering} technique to determine the high-density areas.
Another reason we use clustering is that we can have different sample sizes for
different clusters depending on their \fov distributions. 
In addition, since each cluster may still contain numerous \fovs, we need to
further sample \fovs in each cluster to limit the number of \fovs to be
processed.
We try to sample the minimum fraction of \fovs that still yields accurate
results for the top-k POIs.
To that end, we adopt an \emph{incremental sampling} technique. 
Therefore, we combine both traditional techniques, clustering and incremental
sampling (\cis), into an approach that efficiently detects top-k POIs without
significant loss in accuracy.

%
At the high level, our \cis approach first identifies a set of clusters of \fovs
in the query region, and then incrementally samples \fovs in each cluster. 
For each cluster, a small fraction of \fovs are sampled at each iteration. 
\fovs are sampled without repetitions, and each \fov is processed only once
independetly of the number of sampling iterations.
We know that the more iterations, the more \fovs are considered and the more
accurate the results are, but at the expense of longer processing time.  
Our \cis approach aims to flexibly trade off detection speed and result accuracy
by determining the number of sampling iterations for each cluster with several
heuristic stopping criteria.
These criteria are based on the convergence of the detected top-k POIs and the
capture intention matrix of the culster. 
With this mechanism, the sample size for a cluster of \fovs gradually increases
iteration by iteration approximating to the ``optimal'' sample size. 
Further, the sample sizes of different clusters are decided according to their
own density distributions of \fovs.

Algorithm~\ref{alg:clustering-and-iter-sampling} presents the pseudocode of the
\cis approach. 
We first find the \fovs that overlap with the target region
(line~\ref{cis:range-query-all}) via a range query supported by an R-tree
index~\cite{ay-mm2008,lu-geoinformatica2016}.
Then, we identify $c$ clusters (line~\ref{cis:clustering}) from a uniformly 
random sample (with no repetitions) containing a fraction $f_c$ of the 
\fov population (line~\ref{cis:sampling}). 
Our current implementation uses k-means as the clustering algorithm.
In our experience, $c \in [k, 2 \times k]$ and $f_c \in [0.2, 0.5]$ 
work reasonably well. 
Next, we calculate top-k POIs for each cluster and use them to update the final
top-k POI results 
(lines~\ref{cis:BEGIN-for-each-cluster}--\ref{cis:END-for-each-cluster}).
For each cluster, we obtain all the \fovs $F_c$ that belong to it
(line~\ref{cis:range-query-cluster}). 
\fovs in $F_c$ are then incrementally sampled to update the capture intention
matrix of the cluster 
(lines~\ref{cis:BEGIN-iter-sampling} -- \ref{cis:END-iter-sampling}),
Once the \emph{stopping criteria} are satisfied (line~\ref{cis:END-iter-sampling}),
we obtain from this matrix the top-k POIs identified in the cluster so far. 
The top-k POIs $K_c$ of the cluster are used to update the global 
top-k POIs $K$ (line~\ref{cis:update-global-heap}). 
Since different clusters may have different sample sizes, for fair comparison among
the top-k cells of different clusters, the capture intention value $ci$ of a
cell in a cluster is estimated as $ci/p$, where $p$ is the total fraction of the
cluster's \fov population that was considered (\ie $iter \times f_i$,
line~\ref{cis:pop-fraction}), assuming the clusters have the same sample
size.

\subsubsection{Stopping Criteria} \label{subsubsec:stop-criteria}

The stopping criteria are responsible to tell the algorithm to cut the
iterations short and stop processing a cluster, after having some
\emph{indication} that the top-k cells for the cluster have already been
identified. 
They tend to reduce the detection time, but often at the expense of some loss in
accuracy.

First of all, the iteration number is constrained to not exceed the maximum 
iteration number (\ie $iter \leq 1/f_i$). 
Besides that, we consider two heuristic stopping criteria. 
\begin{compactitem}
\item 
The first criterion monitors changes in capture intention.
It evaluates whether the difference between the maximum capture intentions in
$M_c$ from one iteration to the next is less than a threshold, in which case
the algorithm stops processing the current cluster.
We refer to this criterion as the \emph{difference in maximum capture
intention}. 

\item
The second criterion monitors changes in the locations of the top-k cells from
one iteration to the next.
It calculates the sum of the distances between the closest pairs of cells in
the two top-k result sets from subsequent iterations.
Cells are taken in pairs, one from each result set, in a closest-pair-first
manner, each cell is considered only once, and the distance between their
centers is accumulated. 
The criterion then checks whether such sum is less than a threshold, in which 
case it tells the algorithm to stop processing the current cluster.
We refer to this criterion as the \emph{sum of minimum distances between top-k
results}.
\end{compactitem}

\sref{sec:eval-iter-sampling} examines interesting aspects of the \cis approach,
particularly the trade-offs between detection time and accuracy.

\section{Evaluation} \label{sec:eval}

In this section, we first show the performance gains of the optimized baseline
over the naive baseline.
Then, we evaluate the clustering and incremental sampling (\cis) approach
and study how its parameters influence detection time
and the accuracy of top-k results.

\subsection{Experimental Setup} \label{sec:setup}

\subsubsection{Test Datasets}
We use the datasets in Table~\ref{tab:datasets}. 
They were generated from the GeoUGV dataset\footnote{
Available at \url{http://mediaq.usc.edu/dataset/
}} 
with a Python tool we developed.
GeoUGV~\cite{lu-mmsys2016} is a real-world dataset that includes 2,397 videos
and 208,976 \fovs, collected by $\sim$300 users in more than 20 cities across
the globe between 2007 and 2016.
We modify GeoUGV by varying the maximum visible distance ($R$) and visible angle
($\alpha$) of the \fovs. 
We also augment it with the \fovs{'} upright MBRs for the optimized baseline,
featuring MBR-based cell filtering, to use.
Note that camera's location ($p$) and orientation ($\theta$) remain the same.

\vspace{-5mm}
\begin{table}[h]
\caption{Test Datasets.}
\label{tab:datasets}
\vspace*{-0.07in}
\centering 
\begin{tabular}{|l|l|}
\hline
\bf{Name}  & \bf{Description} \\ 
\hline
DS$_{(100\%,60\degree)}$ & 
  100\% of \fovs with visible angle $\alpha=60\degree$. \\ 
\hline
DS$_{(30\%,160\degree)}$ & 
  30\% of \fovs with $\alpha=160\degree$ and 
  70\% of \fovs with $\alpha=360\degree$. \\ 
\hline
DS$_{(70\%,160\degree)}$  &  
  70\% of \fovs with $\alpha=160\degree$ and 
  30\% of \fovs with $\alpha=360\degree$. \\ 
\hline
\end{tabular}
\end{table}
\vspace{-5mm}


Instead of having a fixed $R$ for all the \fovs (like in GeoUGV), 
we allow $R$ values to vary to make our datasets slightly more realistic. 
We follow the intuition that people more often than not take videos and pictures
of subjects that are close rather than far away, especially in social 
media~\cite{hao-tmm2014}. 
To validate our intuition, we analyzed a sample of videos from the GeoUGV dataset.
We first select all the well-known points of interest (\eg Chinesischer Turm in
Munich,  Marina Bay Sands Skypark in Singapore, the White House in Washington, 
D.C.) in the dataset. Then we search all the videos within a large range 
(say 2km) around each point of interest.
From the sample, we only considered the \fovs of video frames that captured
one of those spots.

\begin{figure}[!tbp]
\centering
\includegraphics[width=0.35\columnwidth]{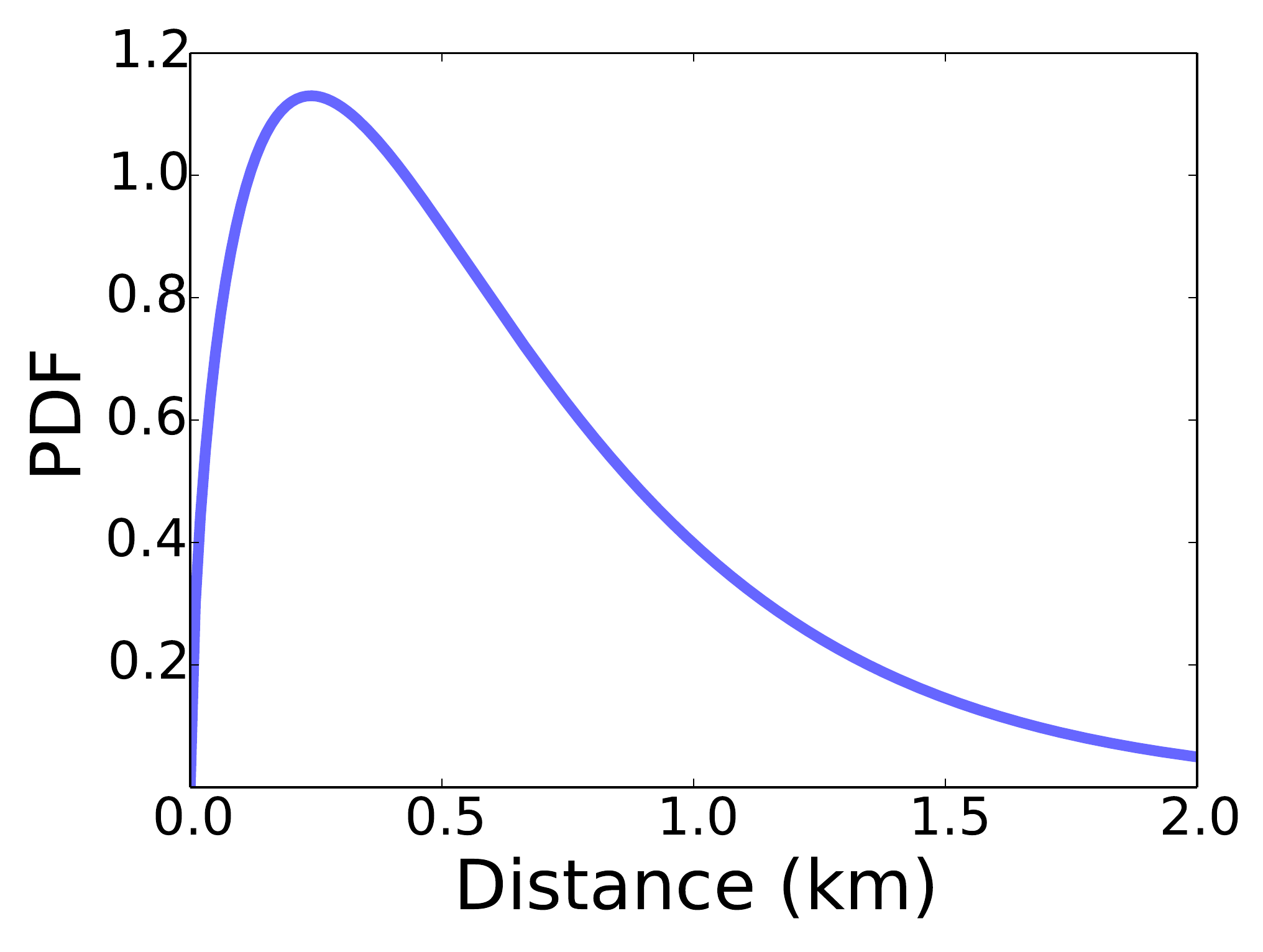}
\vspace*{-0.15in}
\caption{
Probability density function of Gamma distribution (shape=1.6, scale=0.4) used
to generate $R$ values.
}
\label{fig:gamma-distro}
\end{figure}

We calculated the distance between the point of interest and the camera's location
($p$) in over 600 \fovs. 
The results confirmed our intuition: 
videos were taken more often at a closer distance than at a longer distance. 
We also observed that the distance follows a distribution skewed to the right.
For that reason, we generate $R$ values from a Gamma distribution with shape and
scale parameters equal to 1.6 and 0.4, respectively, ensuring that $R<$~2~km
(see \fref{fig:gamma-distro}).
A single $R$ value is obtained per video and assigned to all its \fovs.

We also vary $\alpha$ to simulate video content recorded with $360\degree$
video cameras. 
Samsung Gear 360\footnote{\url{http://www.samsung.com/global/galaxy/gear-360/}} 
and other cameras alike can operate in two modes: 
a dual-lens mode and a single-lens mode. 
We experimented with a Samsung Gear 360 and determined that 
$\alpha = 360\degree$ with both lenses and 
$\alpha \approx 160\degree$ with one lens.
We hence use both angular values in different proportions in our 
datasets DS$_{(30\%,160\degree)}$ and DS$_{(70\%,160\degree)}$.

\begin{table}[!tbp]
\caption{Target regions.}
\label{tab:target-regions}
\vspace*{-0.07in}
\centering 
\begin{tabular}{|c|ccc|}
\hline
\textbf{Locations}  & \textbf{Area} ($\mathbf{km^2}$) & $\mathbf{N_c}$ & $\mathbf{N_f}$   \\
\hline
Merlion Park &     0.78  &       6,446  &  13,483  \\
Munich       &   309.81  &     940,608  &  16,554  \\
Singapore    &   709.96  &   1,441,396  &  64,296  \\
Los Angeles  & 1,297.05  &  12,641,118  &  51,977  \\
\hline
\end{tabular}
\end{table}

\begin{table}[!tbp]
\caption{
Average number of cells covered \newline 
by MBRs of \fovs.
}
\vspace*{-0.07in}
\label{tab:covered-cells-per-fov}
\centering 
\begin{tabular}{|c|ccc|}
\hline
\multirow{2}{*}{\textbf{Locations}} & \multicolumn{3}{c|}{\textbf{Data Sets}} \\ 
        & \scriptsize{$(100\%,60\degree)$} 
        & \scriptsize{$(30\%,160\degree)$}
        & \scriptsize{$(70\%,160\degree)$} \\
\hline 
Merlion Park & 361 & 1,240 &   993  \\
Munich       & 517 & 2,310 & 1,756 \\
Singapore    & 519 & 2,310 & 1,750 \\
Los Angeles  & 519 & 2,314 & 1,771 \\
\hline
\end{tabular}
\end{table}

\subsubsection{Cell Size, Target Regions, and Other Parameters}
We use in the experiments cells of $0.0001\degree$ of latitude by
$0.0001\degree$ of longitude (\ie $\sim$11m$\times$11m).  
Moreover, our evaluation targets the regions in Table~\ref{tab:target-regions};
they have relatively large numbers of \fovs in the GeoUGV dataset, while
differing in area size. 
Note, however, that \fov counts are modest compared to what it is expected in
reality (\eg $>$200 thousand photos and videos uploaded daily to Instagram in
NYC area).
Moreover, Table~\ref{tab:covered-cells-per-fov} lists the average number of
cells covered per \fov in the different regions for our datasets. 

We retrieve from the database all the \fovs covering the target regions in the
time interval from 2010-03-18 to 2016-06-28. 
We verified across the experiments that the query time is just a small fraction
of the total processing time. 
Moreover, the variance parameters for the angular and distance-based capture
intention distributions are $\sigma_{a}=15\degree$ and $\sigma_{d}=25$~m,
respectively, in all the experiments.
These values are suggested in~\cite{zhang-tmm2016} to effectively identify the
points of interest.

\subsubsection{Test Platform and Implementation}
We conduct our experiments on a MacBook Pro laptop running OS~X~10.9.5 and
equipped with 
a 2.6GHz dual-core Intel Core i5-4288U processor, 
8GB of RAM (1600MHz DDR3), and a 512GB SSD. 
We use MySQL Community Server (GPL) v.5.7.15 (with MyISAM engine) to store the
data. 
The table schema is omitted due to space limitations. 

%
%
%

The approaches described in the paper have been implemented in C++11, 
and compiled using \texttt{gcc} with \texttt{-O3} optimization option. 
We use Boost uBLAS library for matrix operations, and \texttt{libmysqlclient}
library to access the database.

\subsection{Optimized Baseline} \label{sec:eval-mbr-cell-filt}

\begin{table}[!tbp]
\caption{
Detection of top-5 cells on DS$_{(100\%,60\degree)}$.
}
\label{tab:cellfiltering-baseline-100pct60degree-top5}
\vspace*{-0.07in}
\centering 
\begin{tabular}{|c|c|cc|}
\hline 
\multirow{2}{*}{\textbf{Locations}} 
  & \textbf{Naive Baseline}
  & \multicolumn{2}{c|}{\textbf{Optimized Baseline}} \\ \cline{2-4}
  & \textbf{Proc. time} & \textbf{Proc. time (Speedup)} & \textbf{Diff.} \\ \cline{1-4}
Merlion Park  & 11.71s              & 2.50s (4.7$\times$)      & 0.0 \\
Munich        & 5,043.92s  (1.4h)   & 8.63s (584.5$\times$)    & 0.0 \\
Singapore     & 30,886.66s (8.6h)   & 30.95s (998.0$\times$)   & 0.0 \\
Los Angeles   & 69,595.80s (19.3h)  & 26.19s (2657.3$\times$)  & 0.0 \\
\hline
\end{tabular}
\end{table}

\begin{table}[!tbp]
\caption{
Detection of top-10 cells on DS$_{(100\%,60\degree)}$.
}
\label{tab:cellfiltering-baseline-100pct60degree-top10}
\vspace*{-0.07in}
\centering 
\begin{tabular}{|c|c|cc|}
\hline 
\multirow{2}{*}{\textbf{Locations}} 
  & \textbf{Naive Baseline}  
  & \multicolumn{2}{c|}{\textbf{Optimized Baseline}} \\ \cline{2-4}
  & \textbf{Proc. time} & \textbf{Proc. time (Speedup)} & \textbf{Diff.} \\ \cline{1-4}
Merlion Park  & 11.72s             &  2.53 (4.6$\times$)       & 0.03m \\
Munich        & 5,044.13s  (1.4h)  &  8.66 (582.5$\times$)     & 0.0 \\
Singapore     & 30,887.22s (8.6h)  & 30.99 (996.7$\times$)     & 0.0 \\
Los Angeles   & 69,601.41s (19.3h) & 26.29 (2647.4$\times$)    & 0.0 \\
\hline
\end{tabular}
\end{table}

\begin{table}[!tbp]
\caption{
Detection of top-5 cells on DS$_{(30\%,160\degree)}$.
}
\label{tab:cellfiltering-baseline-30pct160degree-top5}
\vspace*{-0.07in}
\centering 
\begin{tabular}{|c|c|cc|}
\hline 
\multirow{2}{*}{\textbf{Locations}} 
  & \textbf{Naive Baseline}
  & \multicolumn{2}{c|}{\textbf{Optimized Baseline}} \\ \cline{2-4}
  & \textbf{Proc. time} & \textbf{Proc. time (Speedup)} & \textbf{Diff.} \\ \cline{1-4}
Merlion Park  & 11.71s              & 3.90s (3.00$\times$)    & 0.01m \\
Munich        & 4,975.64s  (1.4 h)  & 25.48s (195.3$\times$)  & 0.0 \\
Singapore     & 31,009.70s (8.6 h)  & 89.23s (347.5$\times$)  & 0.0 \\
Los Angeles   & 61,129.10s (17.0 h) & 75.94s (805.0$\times$)  & 0.0 \\
\hline
\end{tabular}
\end{table}

\begin{table}[!tbp]
\caption{
Detection of top-5 cells on DS$_{(70\%,160\degree)}$.
}
\label{tab:cellfiltering-baseline-70pct160degree-top5}
\vspace*{-0.07in}
\centering 
\begin{tabular}{|c|c|cc|}
\hline 
\multirow{2}{*}{\textbf{Locations}} 
  & \textbf{Naive Baseline}
  & \multicolumn{2}{c|}{\textbf{Optimized Baseline}} \\ \cline{2-4}
  & \textbf{Proc. time} & \textbf{Proc. time (Speedup)} & \textbf{Diff.} \\ \cline{1-4}
Merlion Park & 13.97s              &   5.42s (2.6$\times$)      & 0.0 \\
Munich       & 4,995.01s (1.4 h)   &  31.05s (160.9$\times$)    & 0.0 \\
Singapore    & 31,031.10s (8.6 h)  & 113.44s (273.5$\times$)    & 0.0 \\
Los Angeles  & 56,627.50s (15.7 h) &  94.70s (598.0$\times$)    & 0.0 \\
\hline
\end{tabular}
\end{table}

In this experiment, we evaluate the optimized baseline
(\sref{sec:optimizations}) vs. the naive
baseline (\sref{sec:baseline}).
The performance metrics for comparison are: 
1)~\emph{total processing time}, which includes the time taken by both the
detection procedure and the range query that retrieves the \fovs from the
database; 
and
2)~\emph{difference between top-k results}, measured by the sum of the distances
between the closest pairs of cells in two top-k result sets, as described in
\ref{subsubsec:stop-criteria}.
The latter allows us to measure the accuracy of the optimized baseline compared
to the naive baseline.
Note that our results report average values across 30 runs, except when
individual runs take longer than an hour to complete.

\begin{figure*}[!tbp]
\centering
\subfigure[Naive baseline.]{
\label{fig:heatmap-baseline-merlionpark}
\includegraphics[width=0.30\columnwidth]{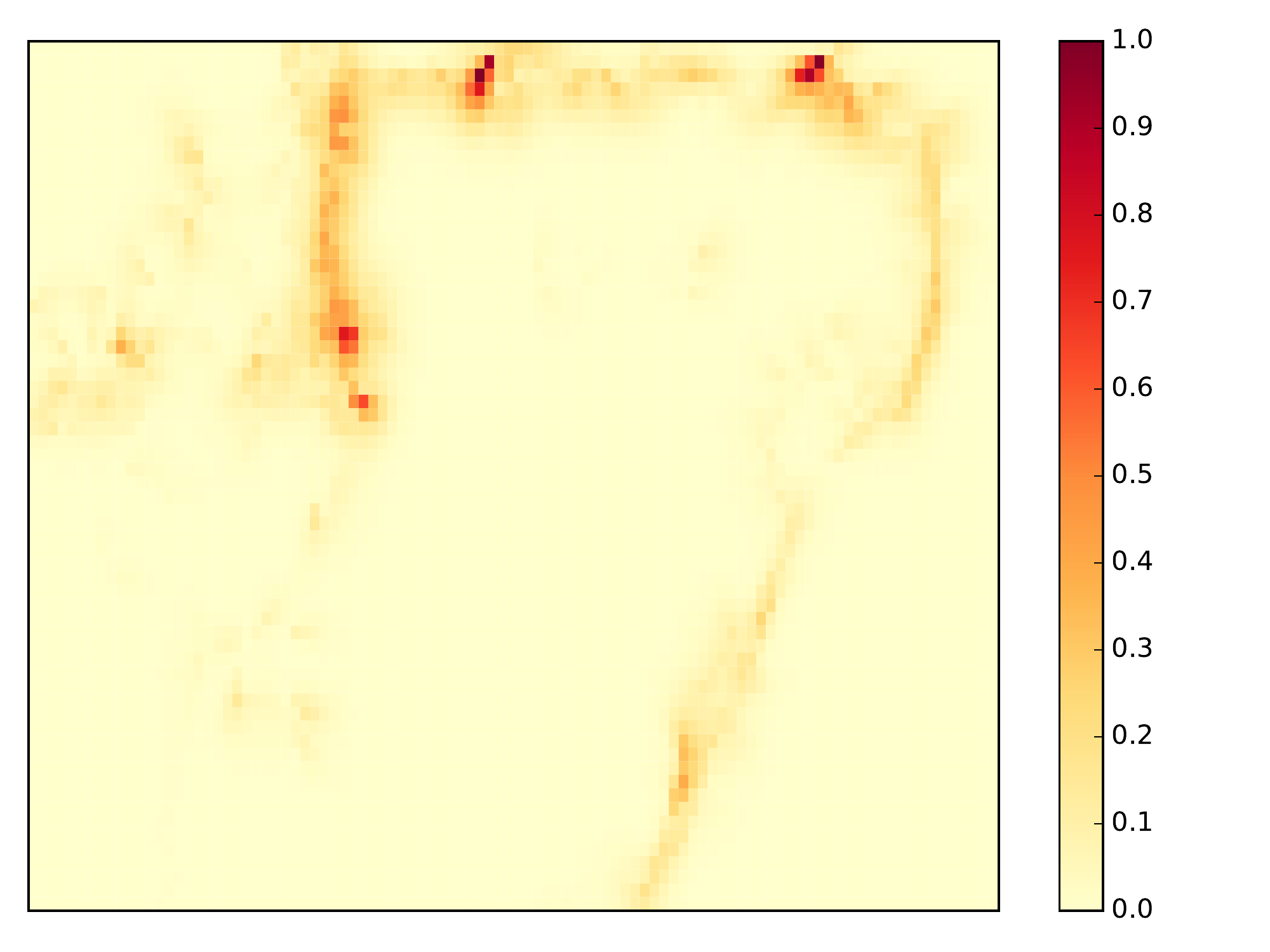}
}
\hspace{0.1in}
\subfigure[Optimized baseline.]{
\label{fig:heatmap-optapproach-merlionpark}
\includegraphics[width=0.30\columnwidth]{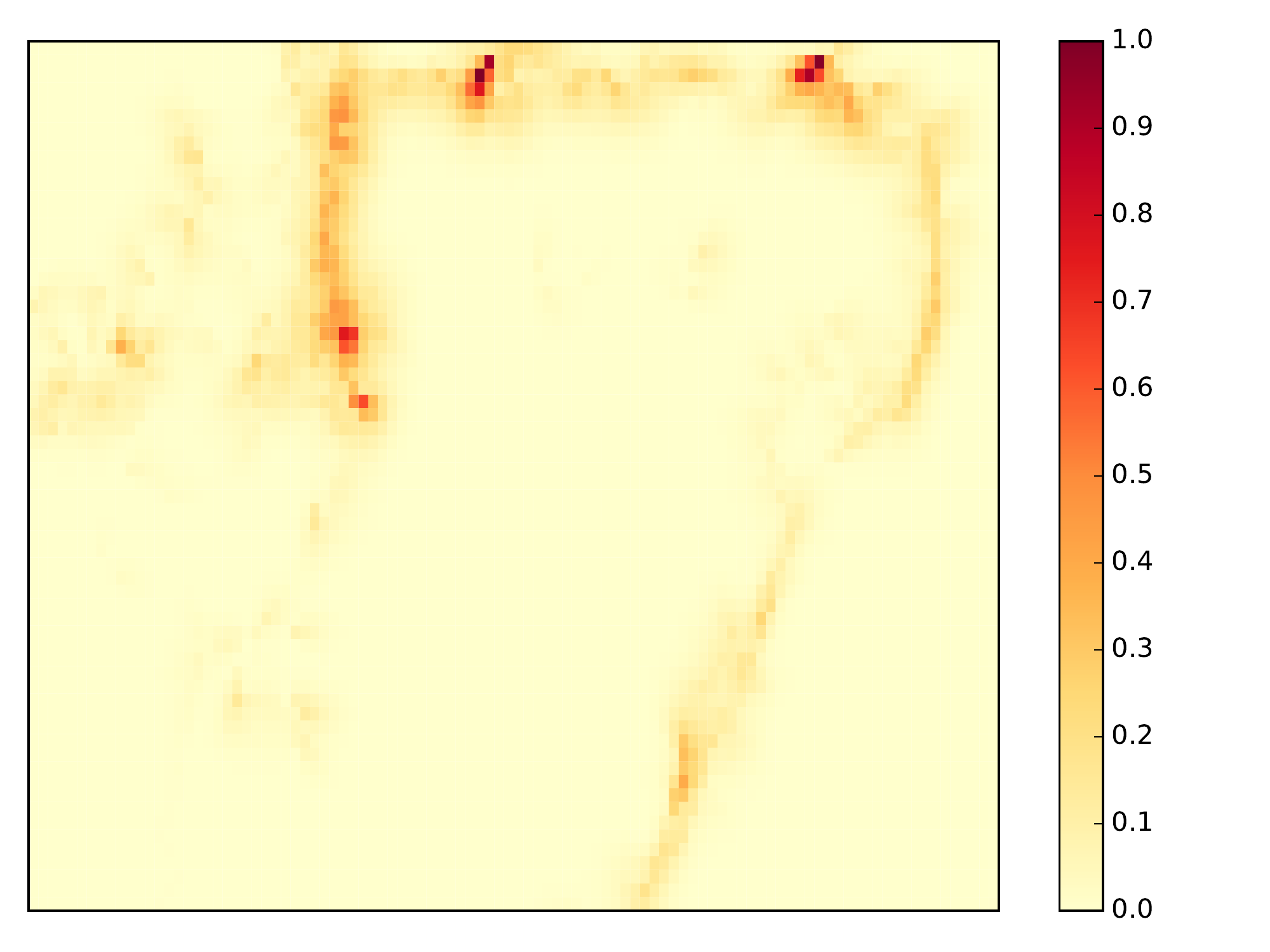}
}
\hspace{0.1in}
\subfigure[\cis approach.]{
\label{fig:heatmap-cisapproach-merlionpark}
\includegraphics[width=0.30\columnwidth]{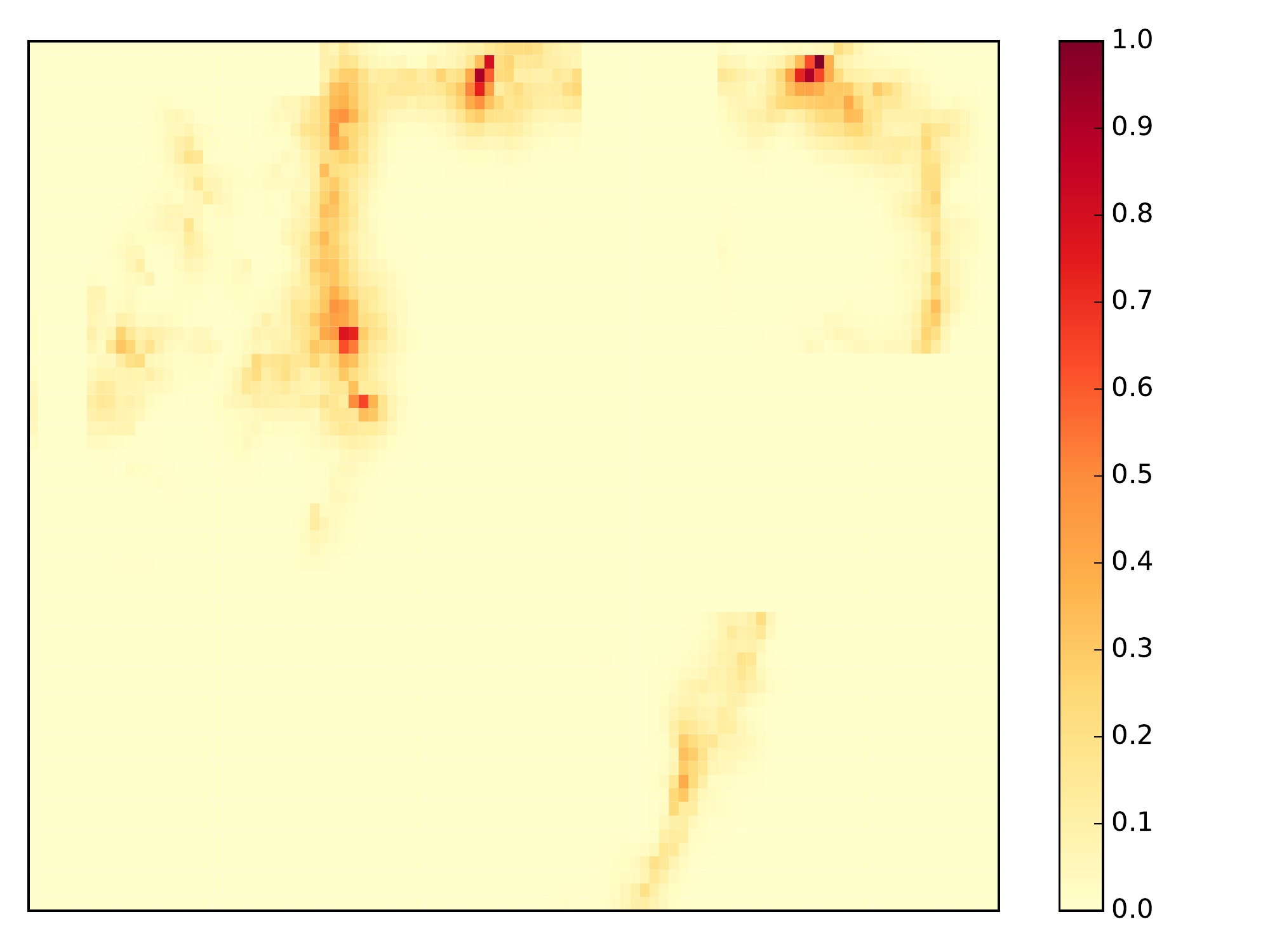}
}
\vspace*{-0.10in}
\caption{
Heatmaps at Merlion Park (Singapore) for DS$_{(100\%,60\degree)}$. 
Although virtually indistinguishable, the heatmaps have some differences,
especially that obtained with the clustering and incremental sampling (\cis)
approach.
}
\vspace*{-0.10in}
\label{fig:cmp-heatmaps}
\end{figure*}


Tables~\ref{tab:cellfiltering-baseline-100pct60degree-top5}
and~\ref{tab:cellfiltering-baseline-100pct60degree-top10} report the average
processing time for the detection of the top-5 and top-10 cells on the
DS$_{(100\%,60\degree)}$ dataset. 
We observe that the optimized baseline, exploiting the MBR-based cell filtering
technique, offers important speedups (4.6$\times$ to more than 2500$\times$)
over the naive baseline with essentially no difference in the top-k results. 
The last point is illustrated in Figures~\ref{fig:heatmap-baseline-merlionpark}
and~\ref{fig:heatmap-optapproach-merlionpark} 
that show identical heatmaps produced by the baseline and the optimized
approaches at Merlion Park.
Similarly, Tables~\ref{tab:cellfiltering-baseline-30pct160degree-top5}
and~\ref{tab:cellfiltering-baseline-70pct160degree-top5} show that, on the
datasets with $360\degree$ visual content, the optimized baseline 
also brings significant reductions (2.6$\times$--805$\times$) in processing time 
while producing the same top-k results.

Our results show that the naive baseline takes very long time to detect top
spots in large areas (\ie Munich, Singapore, and Los Angeles) with modest \fov
counts, compared to those expected in reality.
The reason is that the naive baseline, as explained in \sref{sec:baseline}, uses a
double nested loop to iterate over the \fovs computing the contribution that
each \fov makes to the capture intention of \emph{every} cell in the query area.
By contrast, using MBR-based cell filtering the optimized baseline only
processes the cells that overlap the MBR of each \fov. 
From Tables~\ref{tab:target-regions} and~\ref{tab:covered-cells-per-fov}, we can
see that this represents an important reduction in the number of cells to be
considered per \fov in our test datasets (\eg from 6,446 cells in Merlion Park
to an average of 1,240 or less, and from $\sim$12.6 million cells in Los Angeles
to an average of 2,310 or less). 
Therefore, MBR-based cell filtering, besides being simple and practical, proves
to be very effective in reducing the processing time.

When comparing the results of the optimized baseline across
datasets, we notice that the speedups for DS$_{(30\%,160\degree)}$ and
DS$_{(70\%,160\degree)}$ (\ie the datasets with $360\degree$ visual content) are
smaller than those for DS$_{(100\%,60\degree)}$. 
This is explained again by the differences in the number of cells per \fov in
Table~\ref{tab:covered-cells-per-fov} -- up to 519 in average for
DS$_{(100\%,60\degree)}$, but between $\sim$1,000 and $\sim$2,300 for
DS$_{(30\%,160\degree)}$ and DS$_{(70\%,160\degree)}$. 
Another interesting observation is that the speedup is larger when there are
more circular \fovs with $\alpha = 360\degree$ than with $\alpha = 160\degree$
(\eg 805$\times$ vs. 598$\times$ for Los Angeles), even though circular \fovs
cover more cells. 
This suggests that computing the angular capture intention is a more dominant
factor than the number of cells covered by the \fovs{'} MBRs, 
making the case for the optimization of \sref{sec:impr-4-360-content}.
Finally, 
we have observed that the naive and optimized baselines are both insensitive
to $k$ (the number of top cells being detected). 
The reason is that maintaining a heap with the top-$k$ cells throughout the
detection process is not the dominant factor.


\subsection{Clustering and Incremental Sampling} \label{sec:eval-iter-sampling}

In this section, we evaluate the clustering and incremental sampling (\cis)
approach, presented in \sref{subsec:iter-sampling}.
We use 50\% of the \fov population for cluster
identification ($f_{c}=0.5$), and 5\% of the \fov population in each cluster as
the incremental sample per iteration ($f_{i}=0.05$), for a maximum of 20 iterations
per cluster.
Other parameters, such as cell size, $\sigma_{a}$ and $\sigma_{d}$, remain the
same.
As before, we report average values across 30 runs, unless stated otherwise. 


\begin{figure*}[!tbp]
\centering
\subfigure[Detection time.]{
\label{fig:proctime-vs-clusternum-maxiter-LA}
\includegraphics[width=0.40\columnwidth]{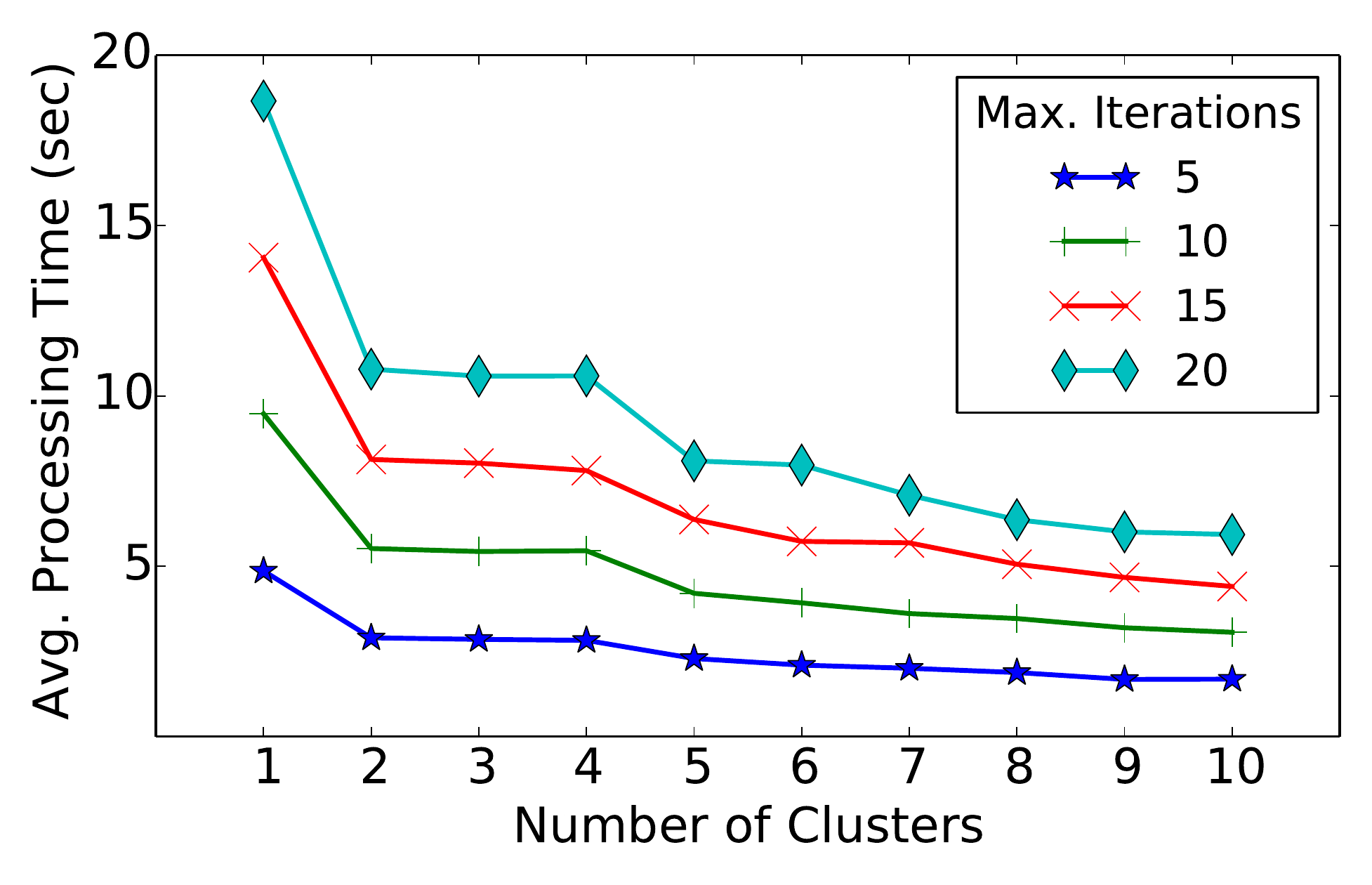}
}
\hspace{3mm}
\subfigure[Total processed cells.]{
\label{fig:proccells-vs-clusternum-maxiter-LA}
\includegraphics[width=0.51\columnwidth]{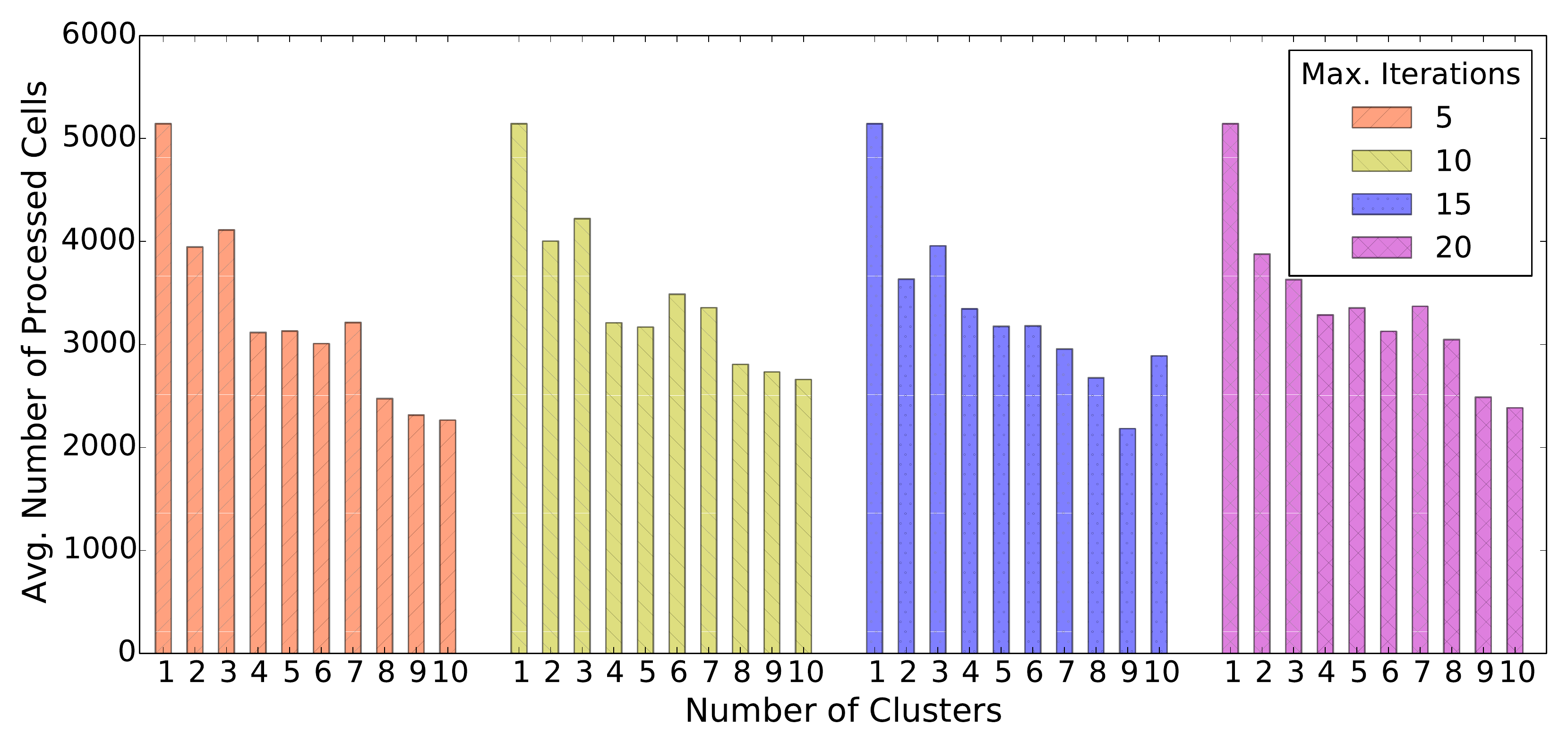}
}
\subfigure[Total processed \fovs.]{
\label{fig:procfovs-vs-clusternum-maxiter-LA}
\includegraphics[width=0.52\columnwidth]{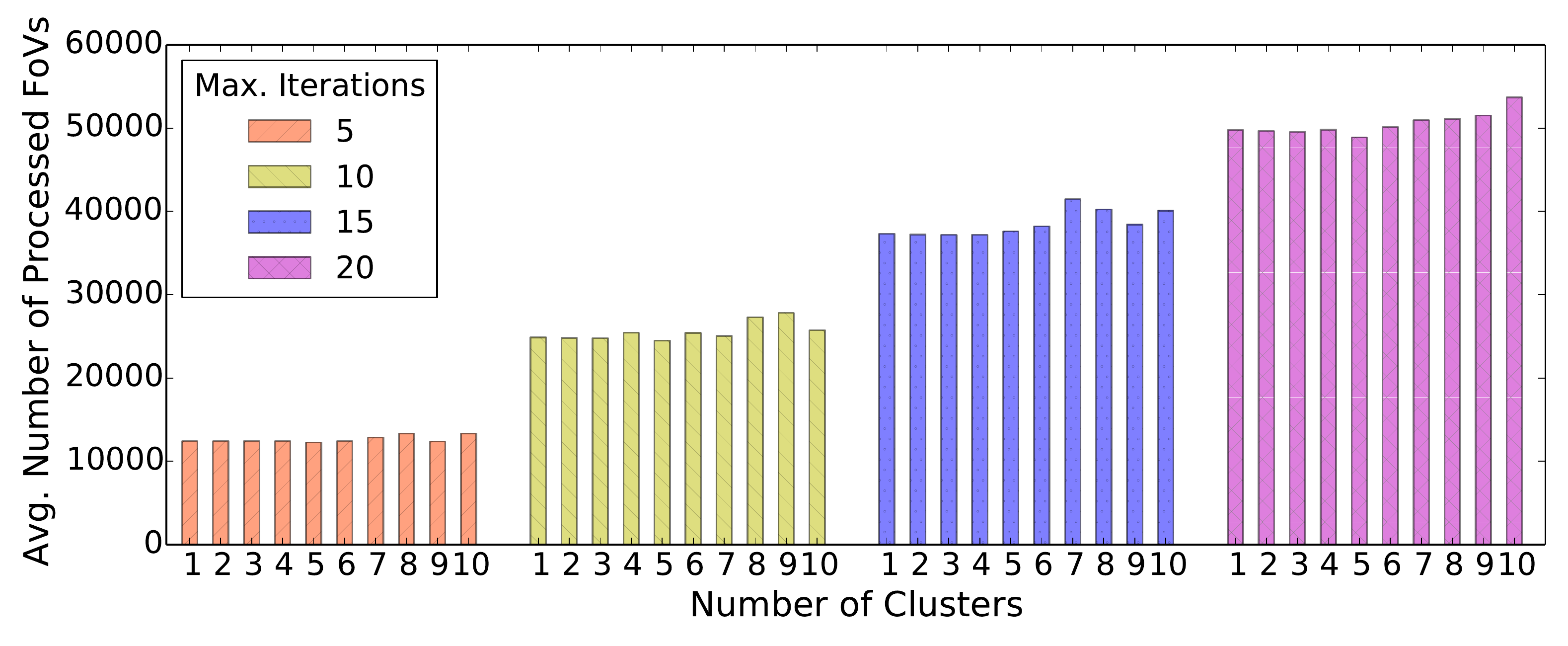}
}
\subfigure[Accuracy]{
\label{fig:distdiff-vs-clusternum-maxiter-LA}
\includegraphics[width=0.45\columnwidth]{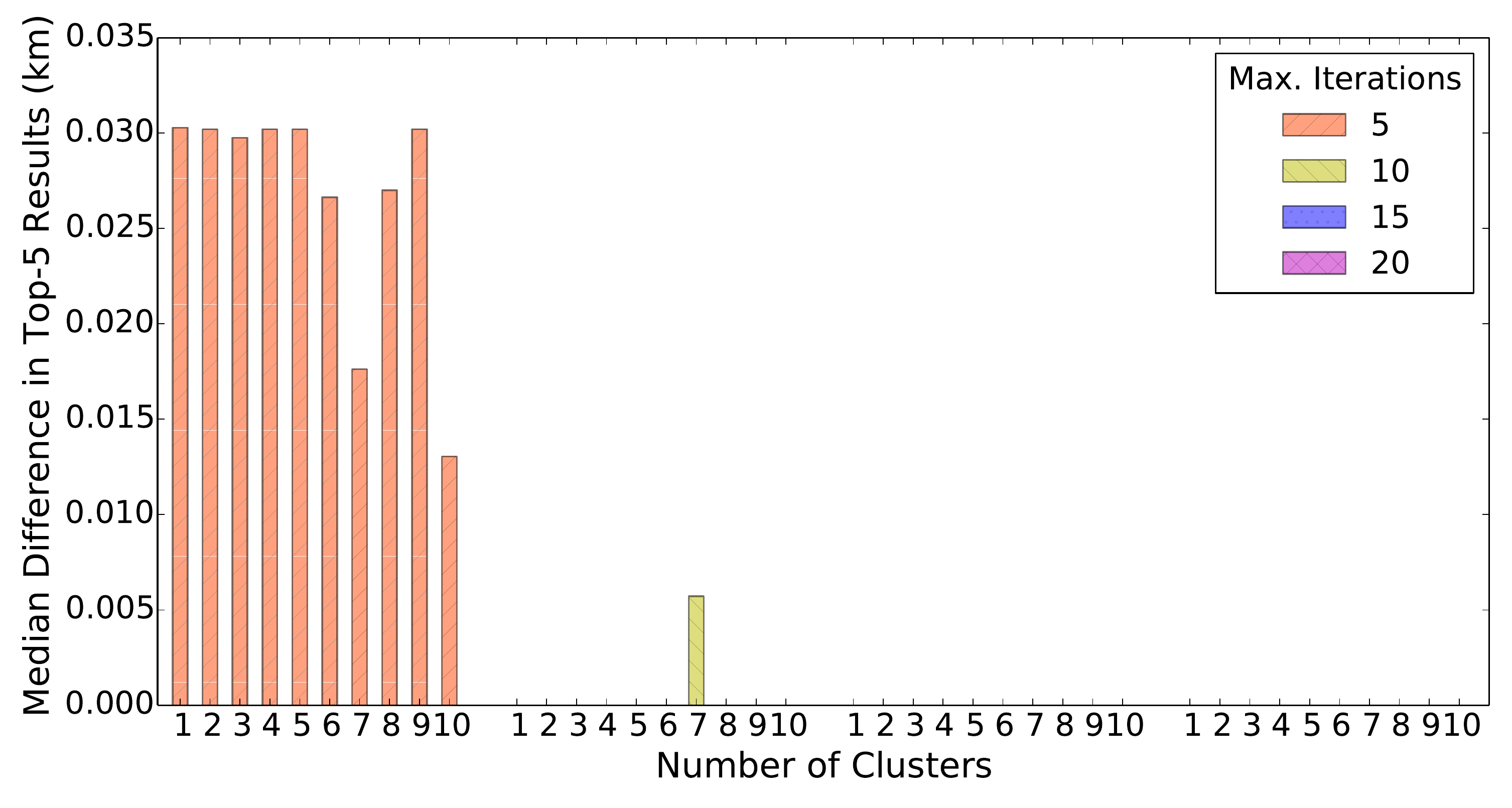}
}
\caption{
Effects of varying cluster counts and maximum iterations in the detection of
top-5 cells in Los Angeles for DS$_{(100\%,60\degree)}$ dataset.
}
\label{fig:vary-clusters-maxiter}
\end{figure*}

We first study how the number of clusters and iterations influence the
processing time and accuracy of the \cis approach on the target
regions for the different test datasets.
We use various cluster counts (from 1 to 10), and stop processing each cluster
at different numbers of iterations (5, 10, 15, and 20).

We observe that the more clusters, the less processing time; this is exemplified
in \fref{fig:proctime-vs-clusternum-maxiter-LA} for Los Angeles area and
DS$_{(100\%,60\degree)}$.
The time reduction is mainly because the number of processed cells decreases as
the number of clusters increases, as shown in
\fref{fig:proccells-vs-clusternum-maxiter-LA}.
By contrast, the number of processed \fovs does not vary greatly with the
cluster count (see \fref{fig:procfovs-vs-clusternum-maxiter-LA}).
Our results then suggest that clustering filters out cells in sparse regions of
the query area and helps focus the detection effort in the most popular regions.  
We also notice that beyond a certain number of clusters (\eg 6 in
\fref{fig:proctime-vs-clusternum-maxiter-LA}) there is no significant additional
gain in detection time. 
In addition, as expected we observe that the more iterations, the longer the
detection time, which is also illustrated in
\fref{fig:proctime-vs-clusternum-maxiter-LA}.
The reason is that the number of processed \fovs increases with the iterations
performed per cluster (see \fref{fig:proccells-vs-clusternum-maxiter-LA}).

More importantly, we notice that in many cases the \cis approach does not need
to reach the maximum number of iterations and process all the \fovs in the
clusters in order to correctly identify the top-k cells.
For example, \fref{fig:distdiff-vs-clusternum-maxiter-LA} shows that it can
obtain the right top-5 cells with just 10 iterations (\ie 50\% of the \fovs in each
cluster).
Consequently, there is opportunity for stopping criteria to cut the iterations short
and identify the top-k cells without much loss in accuracy.


\begin{figure}[!tbp]
\centering
\subfigure[Munich]{
\includegraphics[width=0.35\columnwidth]{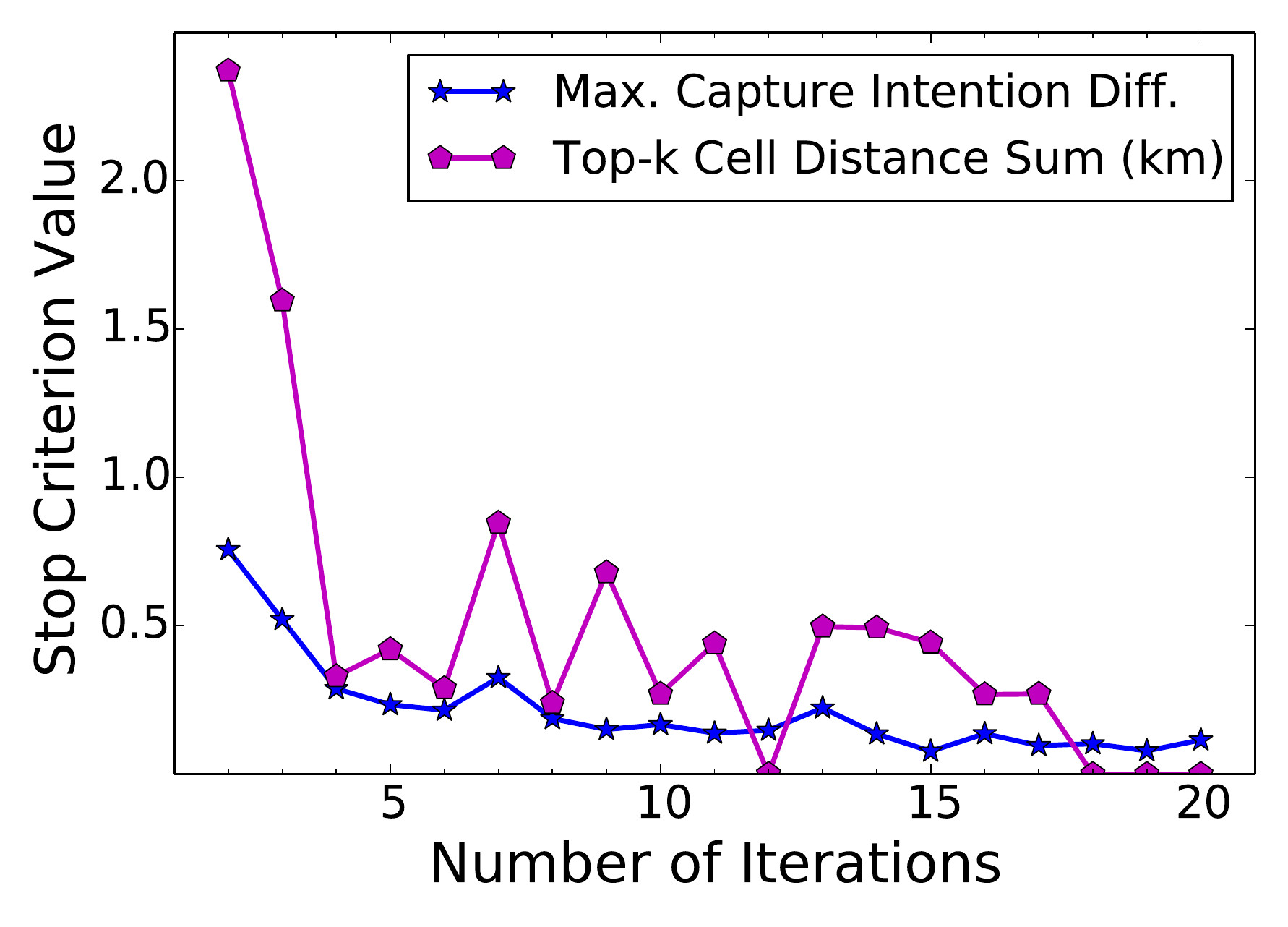}
}
\hspace{4mm}
\subfigure[Los Angeles]{
\includegraphics[width=0.35\columnwidth]{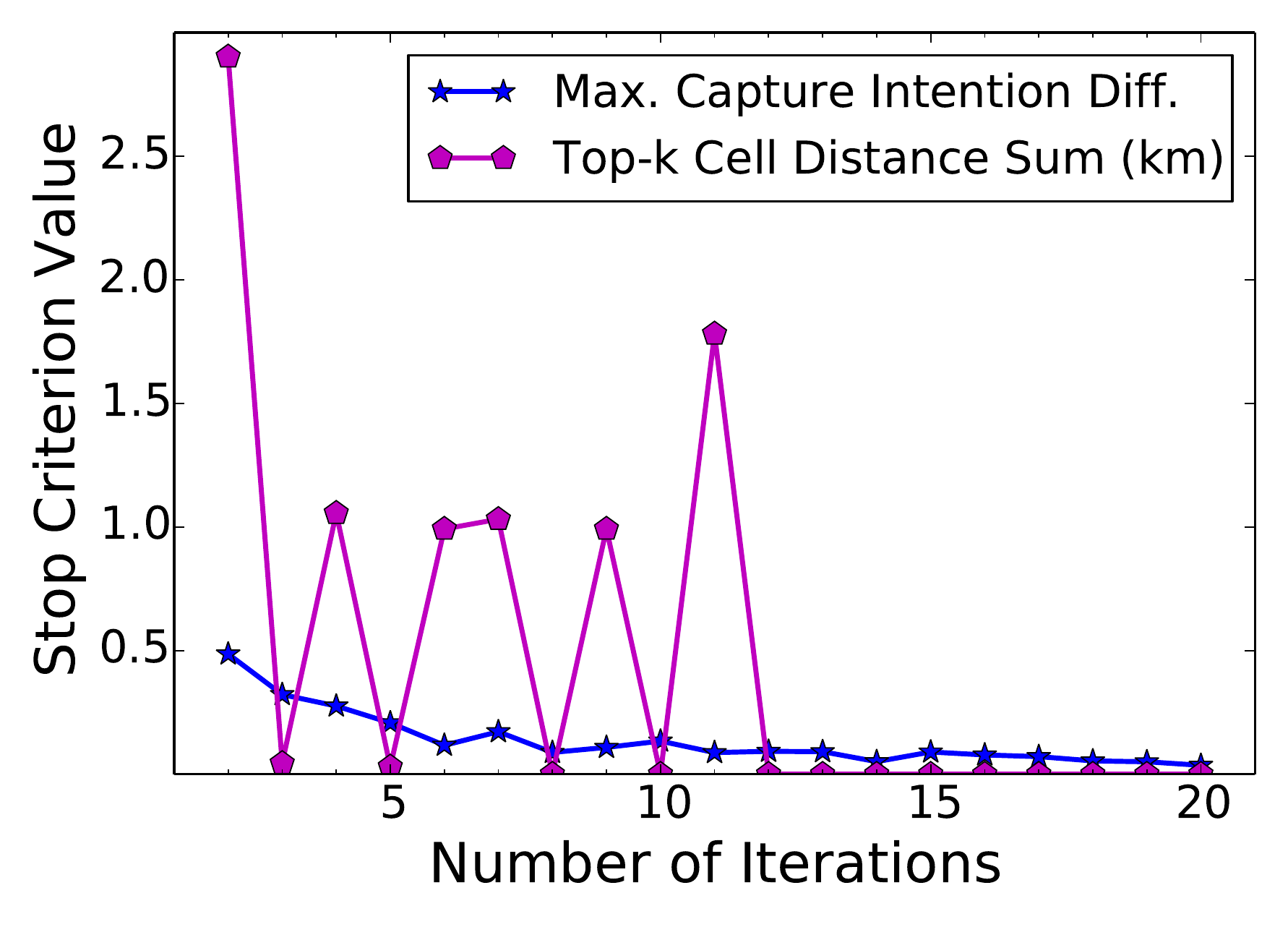}
}
\vspace*{-0.15in}
\caption{
Sample behavior of stopping criteria as iterations progress while processing \fovs
in a cluster for DS$_{(100\%,60\degree)}$.
}
\label{fig:stop-criteria-cmp}
\end{figure}

\begin{figure}[!tbp]
\centering
\subfigure[Detection time with average total number of iterations in
parenthetical values.]{
\includegraphics[width=0.52\columnwidth]{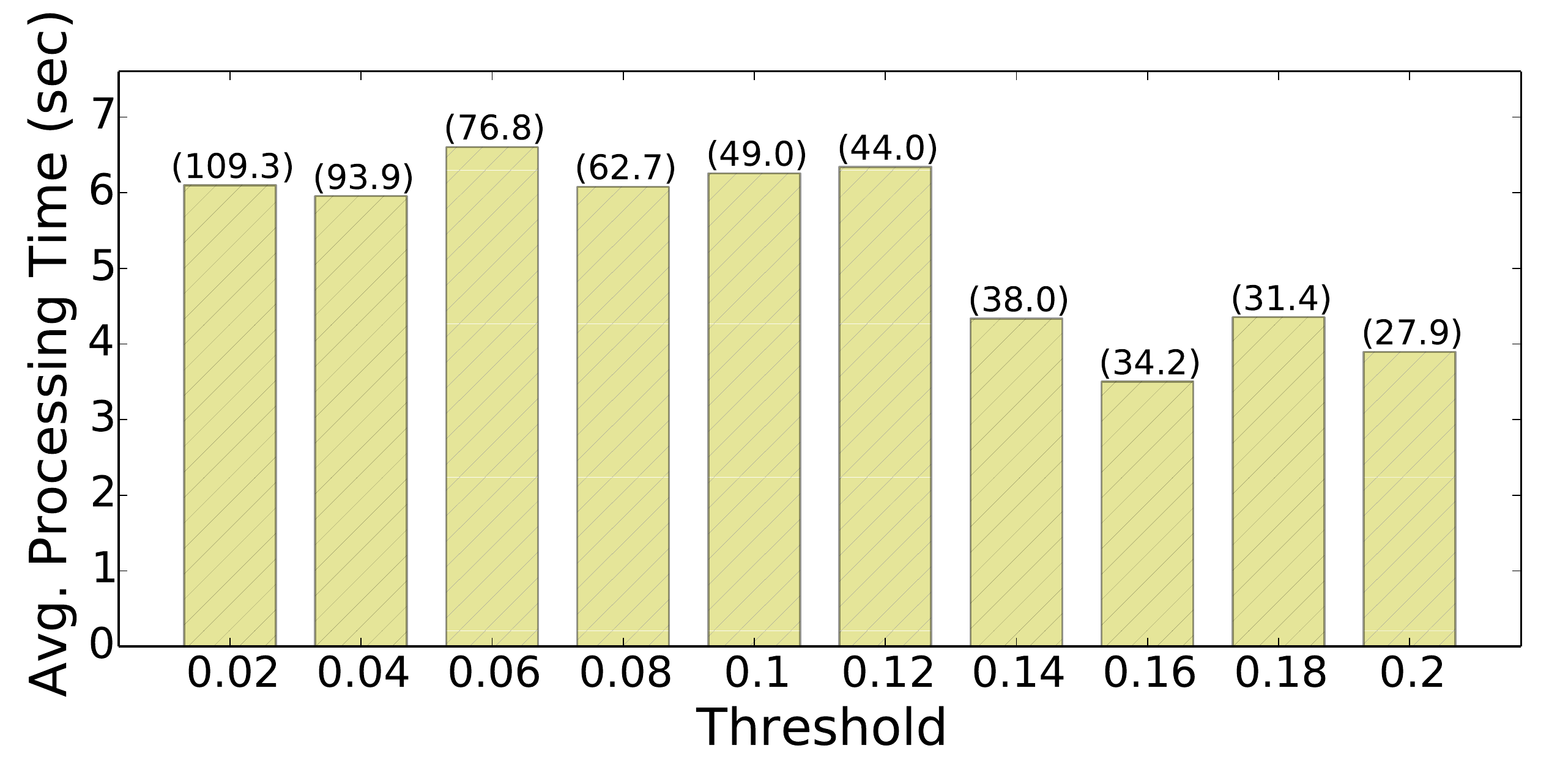}
}
\hspace{2mm}
\subfigure[Loss in accuracy.]{
\includegraphics[width=0.38\columnwidth]{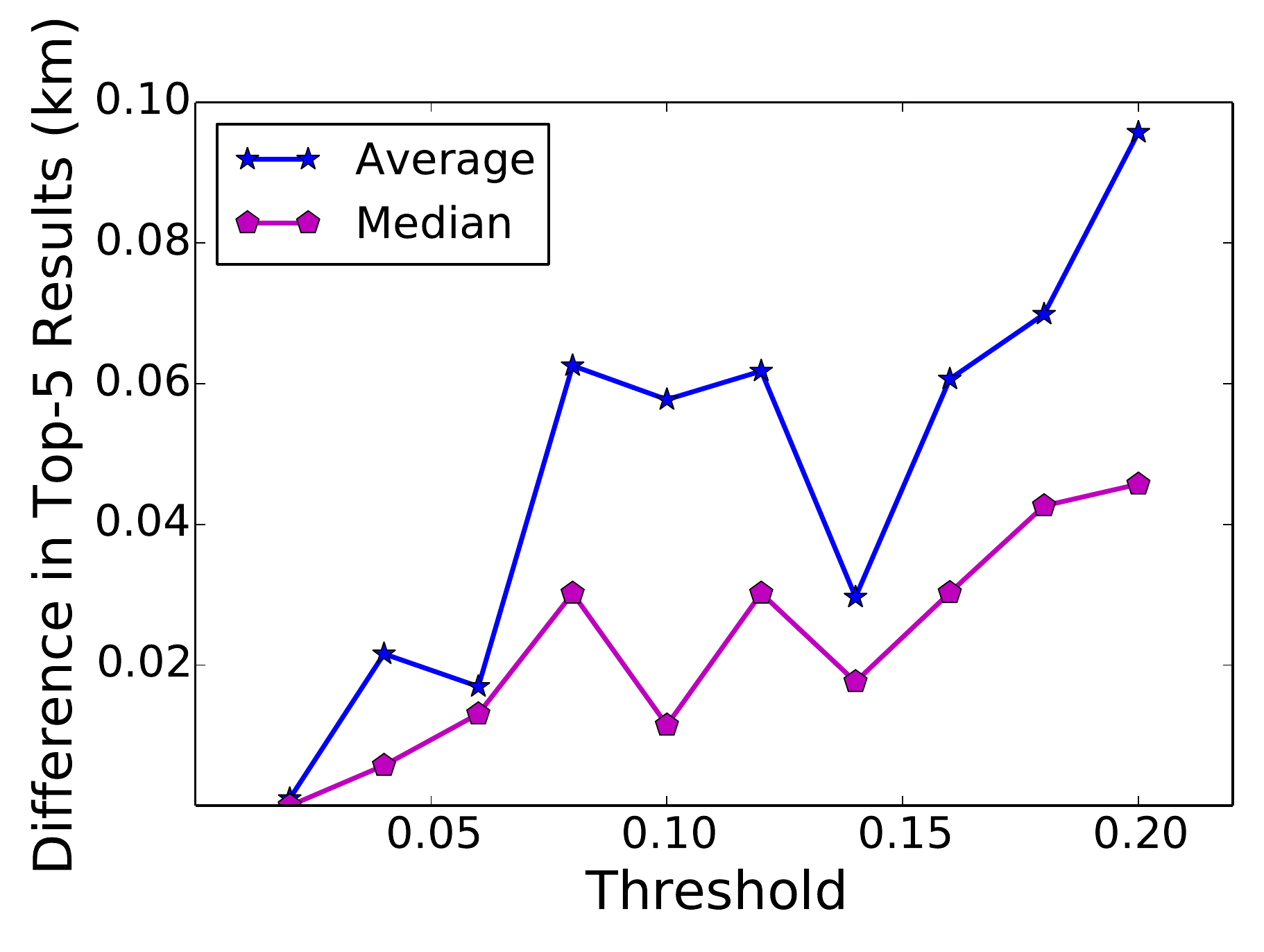}
}
\vspace*{-0.10in}
\caption{
Effect of varying the threshold of the difference in maximum capture intention for Los
Angeles area and DS$_{(100\%,60\degree)}$. 
}
\label{fig:vary-max-ci-diff}
\end{figure}

Next, we consider the stopping criteria introduced in 
\sref{subsubsec:stop-criteria}; they are:
1)~the difference in maximum capture intention, 
and 
2)~the sum of minimum distances between successively identified top-k result
sets.
We evaluate both criteria across the target areas and test datasets, with a
fixed cluster count equal to 6 since it yields reasonably good processing times.
As exemplified in \fref{fig:stop-criteria-cmp}, our results indicate that the
former is more stable, usually decreases and tends to converge as iterations
proceed.
Therefore, the difference in maximum capture intention is the only stop
criterion we use in the rest of this section.  
By further investigating the criterion, we observe that as expected, the higher
the threshold, the shorter the detection time, but the lower the accuracy of the
top-k results.
The reason is that the higher the threshold, the lesser iterations are needed to
satisfy the criterion.
\fref{fig:vary-max-ci-diff} is illustrative of this observation.


We conclude this section with the evaluation of \cis approach's overall
performance for the detection of the top-5 cells in the target regions across
the test datasets.
As before, we use 6 clusters, and we adopt 0.1 as the threshold for the stop
criterion based on the difference in maximum capture intention.
We choose these parameters because they offer just reasonably good detection
time across the datasets, as opposed to highly tuned parameters that yield the
best possible performance for particular cases.
The reason for doing this is that we want to assess how effective and robust the
\cis approach really is in practice.

\begin{table}[!tbp]
\caption{
Detection of top-5 cells on DS$_{(100\%,60\degree)}$.
}
\label{tab:clustering-iterative-sample-cellfiltering-100pct60degree-top5}
\vspace*{-0.07in}
\centering 
\begin{tabular}{|c|cc|ccc|}
\hline 
\multirow{2}{*}{\textbf{Locations}} & \multicolumn{2}{c|}{\textbf{\cis}} & \multicolumn{3}{c|}{\textbf{Single Sampling}} \\ \cline{2-6}
              & \textbf{Time (Speedup)} & \textbf{Diff.}  & \textbf{Sample} & \textbf{Time (Speedup)} & \textbf{Diff.} \\ \cline{1-6}
Merlion Park  & 0.89s (2.8$\times$)     & 251.9m          &  58\%           & 1.53s (1.6$\times$)     &   121.1m    \\
Munich        & 1.51s (5.7$\times$)     &  16.3m          &  30\%           & 2.77s (3.1$\times$)     &    11.5m    \\
Singapore     & 5.88s (5.3$\times$)     & 743.5m          &  20\%           & 6.93s (4.5$\times$)     & 1,536.8m    \\
Los Angeles   & 4.59s (5.7$\times$)     &  30.0m          &  23\%           & 6.41s (4.1$\times$)     &    62.9m    \\
\hline
\end{tabular}
\end{table}

\begin{table}[!tbp]
\caption{
Detection of top-5 cells on DS$_{(30\%,160\degree)}$.
}
\label{tab:clustering-iterative-sample-cellfiltering-30pct160degree-top5}
\vspace*{-0.07in}
\centering 
\begin{tabular}{|c|cc|ccc|}
\hline 
\multirow{2}{*}{\textbf{Locations}} & \multicolumn{2}{c|}{\textbf{\cis}} & \multicolumn{3}{c|}{\textbf{Single Sampling}} \\ \cline{2-6}
              & \textbf{Time (Speedup)} & \textbf{Diff.}  & \textbf{Sample} & \textbf{Time (Speedup)} & \textbf{Diff.} \\ \cline{1-6}
Merlion Park  &  1.17s  (3.3$\times$)   & 1,259.1m        & 66\%            &  2.57s (1.5$\times$)    & 939.4m   \\
Munich        &  1.34s (19.0$\times$)   &     8.4m        & 21\%            &  3.33s (7.7$\times$)    &   4.8m   \\
Singapore     &  6.84s (13.0$\times$)   &     0.5m        & 17\%            & 10.88s (8.2$\times$)    &   0.5m   \\
Los Angeles   &  4.62s (16.4$\times$)   &     8.4m        & 18\%            &  7.90s (9.6$\times$)    &   5.3m   \\
\hline
\end{tabular}
\end{table}

\begin{table}[!tbp]
\caption{
Detection of top-5 cells on DS$_{(70\%,160\degree)}$.
}
\label{tab:clustering-iterative-sample-cellfiltering-70pct160degree-top5}
\vspace*{-0.07in}
\centering 
\begin{tabular}{|c|cc|ccc|}
\hline 
\multirow{2}{*}{\textbf{Locations}} & \multicolumn{2}{c|}{\textbf{\cis}} & \multicolumn{3}{c|}{\textbf{Single Sampling}} \\ \cline{2-6}
              & \textbf{Time (Speedup)} & \textbf{Diff.}  & \textbf{Sample} & \textbf{Time (Speedup)} & \textbf{Diff.} \\ \cline{1-6}
Merlion Park  &  1.37s  (4.0$\times$)   & 1,036.6m        &  76\%           &  3.40s (1.6$\times$)    & 209.0m \\
Munich        &  1.90s (16.3$\times$)   &    16.4m        &  24\%           &  4.18s (7.4$\times$)    &   9.5m \\
Singapore     & 10.61s (10.7$\times$)   &     3.6m        &  22\%           & 14.23s (8.0$\times$)    &   54.2m \\
Los Angeles   &  6.09s (15.6$\times$)   &    22.4m        &  19\%           & 10.07s (9.4$\times$)    &  13.0m \\
\hline
\end{tabular}
\end{table}


Tables 
\ref{tab:clustering-iterative-sample-cellfiltering-100pct60degree-top5}, 
\ref{tab:clustering-iterative-sample-cellfiltering-30pct160degree-top5}, 
and
\ref{tab:clustering-iterative-sample-cellfiltering-70pct160degree-top5}
present the average processing time obtained with the \cis approach.  
The (extra) speedups in the tables are with respect to the optimized baseline's
processing times, reported in 
Tables~\ref{tab:cellfiltering-baseline-100pct60degree-top5},
\ref{tab:cellfiltering-baseline-30pct160degree-top5}
and~\ref{tab:cellfiltering-baseline-70pct160degree-top5}.
As in the previous section, we also report the difference between the
top-k results obtained with the \cis approach and the optimized baseline.
It is calculated as the sum of minimum distances between the two top-k result
sets, which is described in \sref{subsubsec:stop-criteria} and denoted here as
$\sum{d_{min}}$.

Our results show that \cis brings important
($2.8\times$--$19\times$) reductions in processing time over the optimized
baseline, but at the expense of accuracy. 
Even though the top-k results can be reasonably accurate (\ie $\sum{d_{min}} \leq
30~m$), like for Munich and Los Angeles, we see large $\sum{d_{min}}$ values
for Merlion Park with the three test datasets and Singapore with
DS$_{(100\%,60\degree)}$.

As a point of comparison, Tables 
\ref{tab:clustering-iterative-sample-cellfiltering-100pct60degree-top5}, 
\ref{tab:clustering-iterative-sample-cellfiltering-30pct160degree-top5}, 
and
\ref{tab:clustering-iterative-sample-cellfiltering-70pct160degree-top5}
also include results with the \emph{single sampling} approach,
described at the beginning of \sref{sec:single-sampling}. 
To be fair, in each case we configure single sampling to operate on a
\fov sample whose size is equal to the average number of \fovs processed by the
\cis approach -- \ie both approaches use roughly similar fractions of the \fov
population.
We observe that in average \cis is faster than single sampling, and judging from
the $\sum{d_{min}}$ values, both approaches offer somewhat similar accuracy.

\begin{figure*}[tbp]
\centering
\subfigure[DS$_{(100\%,60\degree)}$]{
\includegraphics[width=0.32\columnwidth]{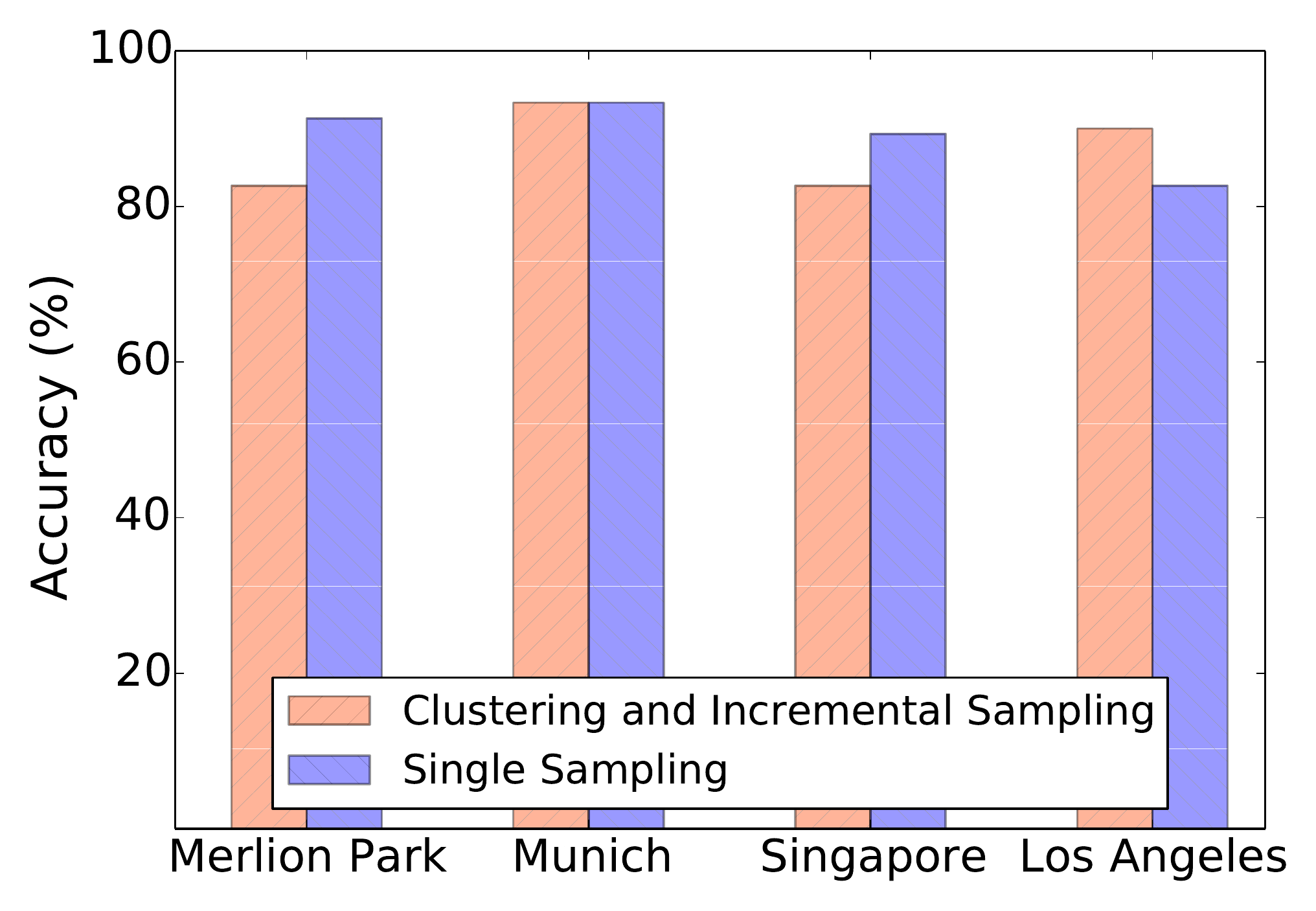}
}
\subfigure[DS$_{(30\%,160\degree)}$]{
\includegraphics[width=0.32\columnwidth]{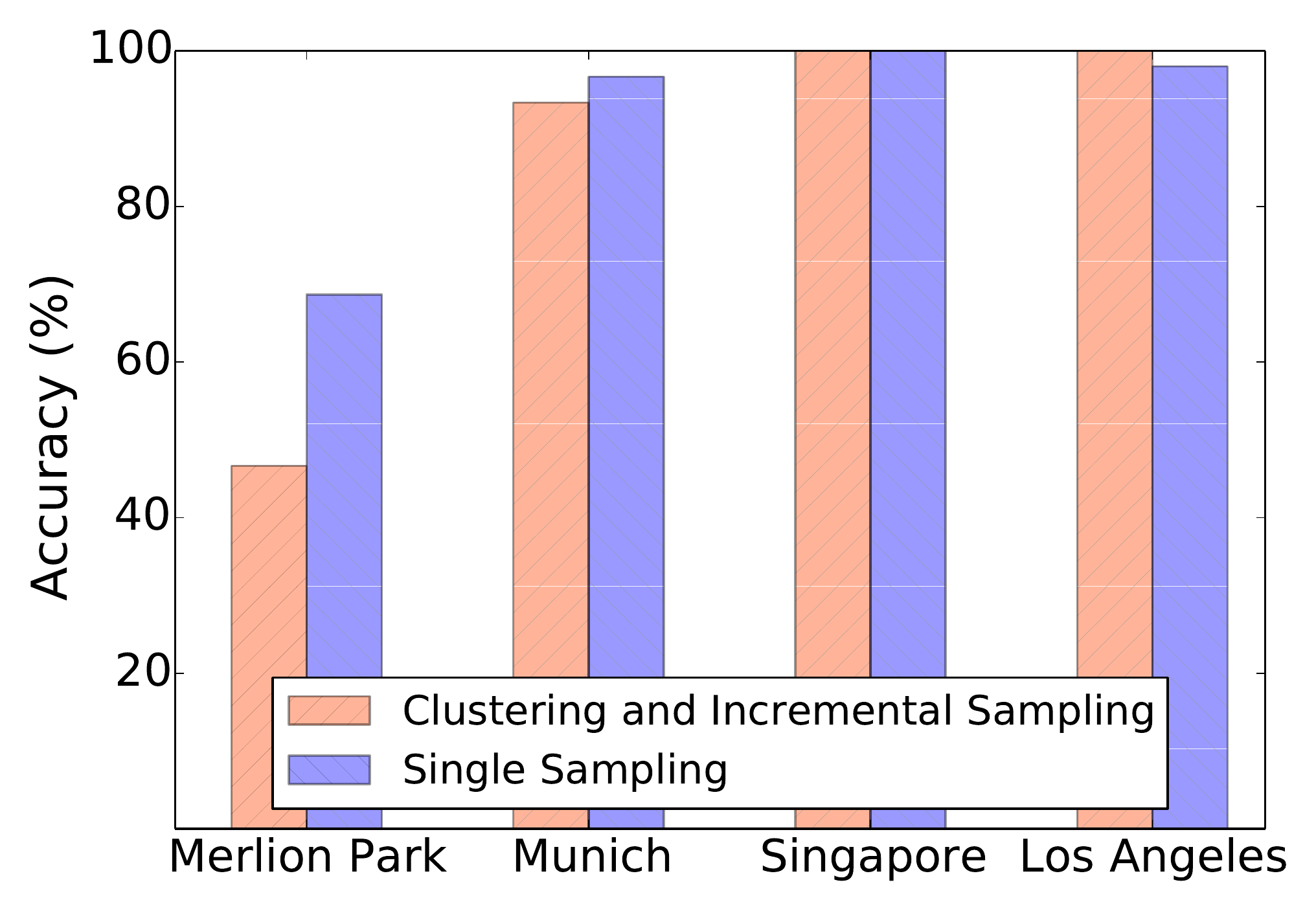}
}
\subfigure[DS$_{(70\%,160\degree)}$]{
\includegraphics[width=0.315\columnwidth]{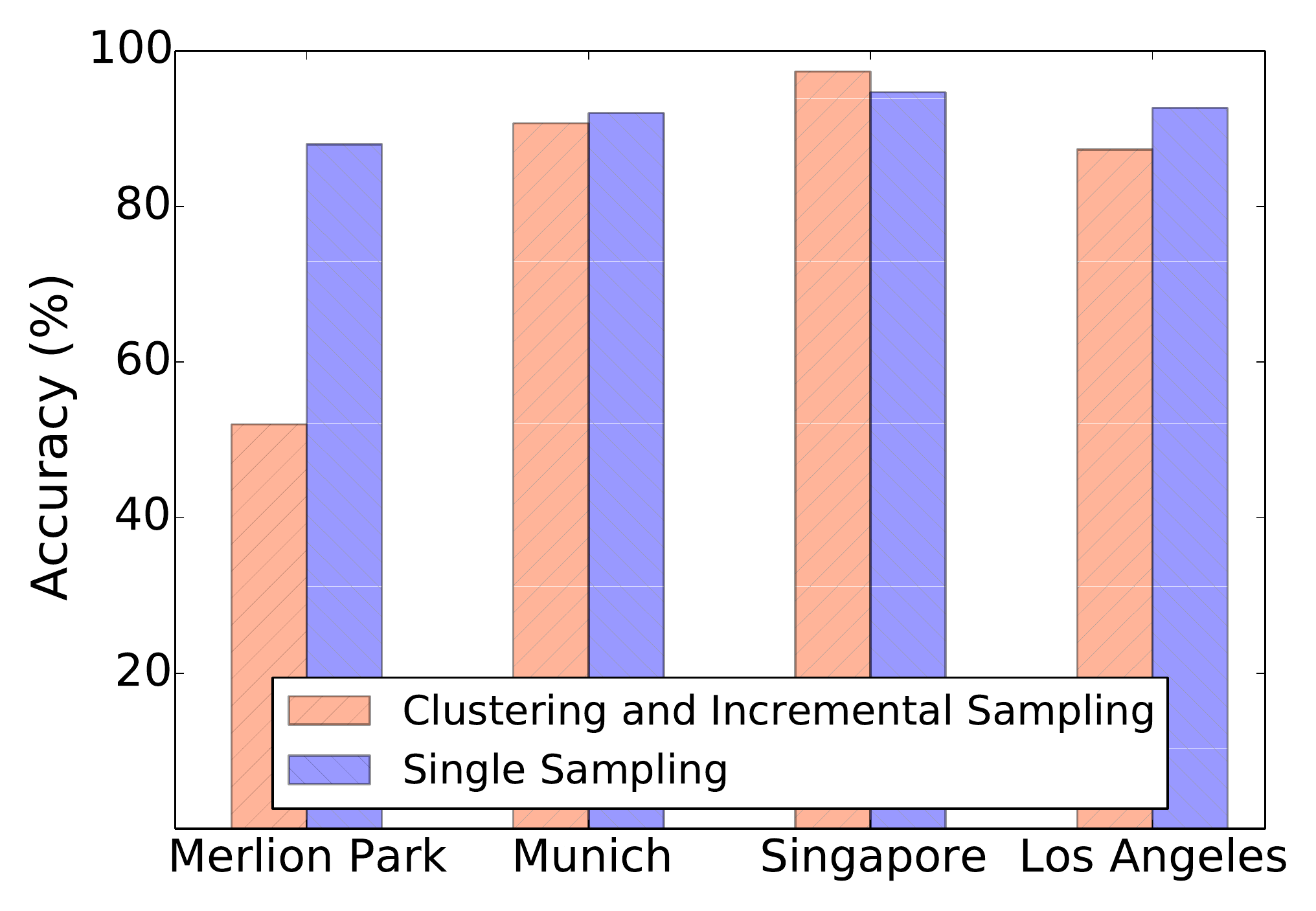}
}
\vspace*{-0.10in}
\caption{
Percentage of top-5 cells correctly identified by the clustering and
incremental sampling approach and the single sampling approach.
}
\label{fig:accuracy-cmp}
\end{figure*}

We further investigate the accuracy of \cis and single sampling approaches by
examining the percentage of top-5 cells they correctly identify, compared to
those obtained with the baseline.
We consider that a cell has been ``correctly'' identified if it is 20~m or less
from the closest actual top cell (\ie it is adjacent to the actual top cell).
Here, cells also are taken in pairs, one from each top-k result set, in a
closest-pair-first manner, and each cell is considered only once.

\fref{fig:accuracy-cmp} indicates that in most cases both \cis and single
sampling detect 
at least 80\% of the top-k cells (\ie 4 out of 5).
For example, \fref{fig:heatmap-cisapproach-merlionpark} shows a heatmap produced
by \cis, on Merlion Park for DS$_{(100\%,60\degree)}$, that is virtually
indistinguishable to the naked eye from the heatmaps in
Figures~\ref{fig:heatmap-baseline-merlionpark} and
\ref{fig:heatmap-optapproach-merlionpark} by the baseline and optimized
approaches; although there is some difference. 
Nevertheless, it is precisely at Merlion Park, but with DS$_{(30\%,160\degree)}$
and DS$_{(70\%,160\degree)}$ (\ie $360\degree$ visual content), where the \cis
approach performs rather poorly with the current parameters.


%
%
%
%

\section{Discussion} \label{sec:disc}

We have shown that the optimized baseline offers significant reductions in
detection time over the naive baseline, with virtually no loss in accuracy. 
It incorporates practical optimizations that have proven very effective. 
Particularly, MBR-based cell filtering makes a strong case for
storing the \fovs along with their upright minimum bounding rectangles. 

We have also shown that the \cis approach (\sref{subsec:iter-sampling}) is able to
reduce detection time even further, but by sacrificing the accuracy of the
results.
\cis was effective in a number cases (\eg Munich and Los Angeles).
But in other cases (\eg Merlion Park), the accuracy loss was noticeable in terms
of distance.  
This is because the popular subjects at different locations in Merlion Park have
comparable capture intentions, as calculated from the GeoUGV dataset (see
\fref{fig:heatmap-baseline-merlionpark}). 
Due to the use of random sampling, the \cis approach may miss popular POIs in
areas where other POIs exist with similarly high capture intentions.
This observation suggests that \cis is more suitable for detecting 
POIs whose capture intentions are more diverse.


According to our evaluation, the main reason for accuracy loss is incremental
sampling, which has two main components: the \fov sampling method, and the
stopping criterion.
First, we draw samples of the \fov population uniformly at random (without
repetitions).
Besides its simplicity and being a popular choice, this method was chosen
because it sets a reference point for more sophisticated methods.  
We expect that more advanced sampling methods will better guide the
incremental selection of \fovs and help cap the accuracy loss. 
Second, the adopted stopping criterion based on the difference of maximum capture
intention is rather simple and in some cases, clearly insufficient to ensure
good accuracy. 
Other more elaborate stopping criteria may be more effective.

In Merlion Park, we also observed that the accuracy of the results improved by
reducing the fraction of circular \fovs ($\alpha=360\degree $) in the dataset.
That suggests that the angular capture intention $ci_{a} \triangleq 1$ of
circular \fovs, which makes cells less differentiable, has some negative effect
when, rather than all, only a sample of \fovs is considered.

Note that using the optimized baseline on small regions (like Merlion Park) and
the \cis approach on large ones is also a practical alternative.

\section{Related Work} \label{sec:relwork}

Generally speaking, previous work focuses on two problems:
1)~detection of points or regions of interests, and
2)~identification and retrieval of the top-k of those spots of interest. 
Below we summarize research efforts in those areas most relevant to our work.

\subsection{Detection of Points and Regions of Interest}

Some approaches \eg~\cite{duygulu-eccv2002,sivic-iccv2005})
identify POIs from visual content by extracting and analyzing image features. 
For example, Duygulu et al. use content-based techniques to extract image features
which are then matched to keywords taken from bag-of-words vocabularies.
They are computationally intensive and thus not applicable to large volumes of
visual content. 
By contrast, instead of analyzing the visual content, other approaches
(\eg~\cite{liu-neurocomputing2013,zheng-cvpr2009,hao-tmm2014,zhang-tmm2016}) 
accelerate the detection
of interesting locations or objects by leveraging sensor-generated metadata (\eg
GPS locations) or keyword tags associated with the visual content. 
The approaches presented in this paper belongs to this group.
For example, Liu et al.~\cite{liu-neurocomputing2013} propose a filter-refinement 
framework to discover hot topics based on the spatio-temporal distributions of 
geo-tagged videos from YouTube. 
Zheng et al.~\cite{zheng-cvpr2009} built a landmark recognition engine 
which models and identify the landmarks automatically from geotagged 
photos at the world scale.
Unlike our work that considers \fovs, these frameworks only use the camera 
locations to describe the visual contents whereas the locations that are 
of interest to people may be far away.

The two recent studies~\cite{hao-tmm2014,zhang-tmm2016} are the most closely related 
to our work. Hao et al.~\cite{hao-mm2011,hao-tmm2014} represent each video frame as a 
camera view (\ie a vector pointing along the camera shooting direction)
and propose two methods to detect POIs: 
1)~a cluster-based method and 
2)~a grid-based method.
The cluster-based method computes the intersection points of all the camera views 
and from these intersections, infers clouds of points as POIs.
The grid-based method, on the other hand, divides the space into grid cells, 
generates a heatmap based on how often a cell appears in different camera views,
and then identifies the popular the places. 
A sector-based cell filtering technique is applied to accelerate the detection. 
However, this study processes all the \fovs, which is still not efficient for large-scale
\fovs.
Further authors in~\cite{hao-mm2011,hao-tmm2014} assume that people's intention
is to only capture targets at the center of the scene (\ie aligned with the
camera's shooting direction).
However, targets may be located in different places withing the visible area.
To overcome this limitation, Zhang et al.~\cite{zhang-tmm2016} 
propose an \fov-based approach that applies a probabilistic model to describe
people's capture intention (see \sref{sec:baseline}). 
They experimentally show that their approach offers much higher accuracy than
the approaches in~\cite{hao-mm2011,hao-tmm2014}. 
For that reason, we adopted this approach as our baseline, and have shown that
our improvements offer significant speedup.

Other efforts try to detect points or regions of interest from other data
sources. 
For example, Vu et al.~\cite{vu-georich2016} present a framework for estimating 
social point-of-interest boundaries from spatio-temporal information in 
geo-tagged tweets. 
Ye et al.~\cite{ye-sigir2011} and Yuan et al.~\cite{yuan-sigir2013}
provide POI recommendation approaches based on users' check-in behaviors.
Gao et al.~\cite{gao-aaai2015} build a content-aware POI recommendation system
by relating the content information on location-based social networks (i.e., 
POI properties, user interests and sentiment indications) 
to check-in actions.

Note that most of the works mentioned above focus on identifying all the points
of interests in an area. Our work, however, focuses on top-k detection.

\subsection{Retreival of Top-k Points of Interests and Objects}

Peng et al.~\cite{peng-sigir2013} propose a probabilistic field of view (pFoV)
model for smartphone photos to capture the uncertainty in camera sensor data. 
Given a database of POIs, a set of geotagged photos represented in the pFOV model, and 
a query photo, authors identify the most prominent POI captured in the query photo.  
Skovsgaard et al.~\cite{bogh-pvldb2013,skovsgaard-gis2014} focus on retrieving
the top-k points of interest from spatial-keyword data (\eg geo-tagged twitter
data).
Toyama et al.~\cite{toyama-mm2003} introduce an image database that supports
image indexing and search based on the image camera locations and recording
time.

The research work mentioned above assume that the points of interest are given,
whereas in this paper we focus on detecting those spots without prior knowledge.

\section{Conclusions} \label{sec:conc}

In this paper, we have presented an efficient approach to detecting top-k points of
interest from geotagged visual content in a user-specified area. 
Based on clustering and incremental sampling, it trades off accuracy of top-k
results for detection speed.
We provided a thorough evaluation of the speedups as well as accuracy losses of
the proposed approach. 
Our results show that the \cis approach offers 2.8$\times$--19$\times$
reductions in processing time over an optimized baseline, while in most cases
correctly identifying 4 out of 5 top locations.

We also introduced two simple, yet effective optimization techniques that enable
significant reduction in detection time with no accuracy loss.
In particular, the MBR-based cell filtering technique makes a strong case for
storing \fovs along with their MBRs.

In the future, we plan to study advanced sampling methods and more sophisticated
stopping criteria that could offer accuracy improvements. 
Moreover, despite the obtained speedups, our prototype implementation could be
optimized even further.
For example, we could leverage multicore processors and process the clusters in
parallel. 
Caching techniques could also be used to reduce response time since
multiple users are likely to make the same or similar queries.


\section*{Acknowledgments}

We thank our colleagues in the Smart Systems team at Samsung Research America
for their feedback.  
Ying Lu's work was partially funded by 
the NSF grants IIS-1320149, CNS-1461963, and the USC Integrated Media Systems Center.  
Except for funding, neither sponsor contributed to or influenced any part of
this paper.  
Nothing herein represents the views and opinions of the sponsors.

\small
\bibliographystyle{splncs}
\bibliography{main}

\end{document}